\documentclass[aps,prd,twocolumn,groupedaddress]{revtex4-1}

\usepackage{graphics,graphicx}
\usepackage{amsmath, amssymb}
\usepackage{natbib,hyperref}
\usepackage{mathtools}

\begin{document}

\title{Density of States of a Coupled-Channel System}

\author{Pok Man \surname{Lo}}
\email[]{pokman.lo@uwr.edu.pl}
\affiliation{University of Wroc\l aw, PL-50204 Wroc\l aw, Poland}

\date{\today}

\begin{abstract}
We demonstrate how an effective density of states can be derived from the S-matrix describing a coupled-channel system.
Besides the locations of poles, the phase of the determinant of the S-matrix encodes essential details in 
characterizing the dynamics of resonant and non-resonant interactions.
The density of states is computed for the two channel scattering problem ($\pi\pi, K \bar{K}$, S-wave), 
and the influences from the various dynamical structures: poles, roots, branch cuts, and Riemann sheets, are examined.
\end{abstract}

\maketitle

\section{Introduction}

Thermodynamics is essentially tied to the proper counting of states: 
The question is what states to count and how to count them.
In the scattering matrix (S-matrix) formulation of statistical mechanics, 
the density of states (DoS), i.e. the gradation of the number of states in energy,
is expressed in terms of the S-matrix~\cite{dmb,prakash}:

\begin{align}
	B(E) =  \frac{1}{2} \, {{\rm Im} \, \rm Tr} \, \left[ \, S^{-1} \frac{\partial}{\partial E}  S - \left(\frac{\partial}{\partial E} S^{-1} \right) S \, \right].
\end{align}
The partition function is then given by an integral of the DoS 
with the appropriate Boltzmann weight. 
A unique feature of this formulation, in contrast to the standard Matsubara approach,
is the decoupling of zero temperature dynamics and statistics~\cite{dmb,leclair1,smat}.
This is what makes the scheme powerful: 
one can make progress in understanding the thermal medium by successively improving the S-matrix input: 
including relevant channels, extension to $N>2$ scatterings, etc., 
thus working towards building an accurate virial / cluster expansion.

There are clear advantages in writing the DoS in terms of the S-matrix:
The S-matrix has a direct connection to (existing and future) experimental data~\cite{pdg}, 
accompanied by powerful theoretical tools such as 
chiral perturbation theory~\cite{Gasser:1983yg, Oller}, 
lattice QCD~\cite{Shepherd:2016dni}, effective hadron models~\cite{Rapp:1999ej} and potential models~\cite{Godfrey:1985xj,Barnes1992}. 
An active research program has begun to leverage the very precise information on particle spectra 
in studying the thermal properties of hadron systems, 
such as those created by heavy ion collisions~\cite{Broniowski_2003,Ortega:2017hpw,Lacroix2015,rho,omega,ppuzz} or in astrophysical systems~\cite{Horowitz:2005zv,Horowitz:2005nd,Mallik:2008zt,Ropke:2012qv,Oertel:2016bki,sanjay};
and for interpreting LQCD results at finite temperatures~\cite{Borsanyi:2011sw, Bazavov:2012jq,Huovinen:2017ogf,kmat,kappa,chibq,chibs}.

The connection to the S-matrix also brings in an interesting theoretical issue:
How a degree of freedom (DoF) is represented by the open channels, on which the S-matrix acts on.
In the conventional construction of a Hamiltonian (or a Lagrangian), 
one often needs to distinguish composite particles from elementary particles. 
In the S-matrix framework, the (assumed asymptotically complete) Hilbert space is spanned by the scattering states (plus bound states)~\cite{taylor},
this makes it necessary to understand how resonances, and other dynamically generated states, 
are incorporated in the scheme.

For a narrow resonance in a single (2-body) scattering channel, there is a clear answer:
the presence of the resonance is encoded in the phase shift, 
the latter behaves like a step function.
The contribution to the thermal trace is as if the resonance is a point-like, 
elementary particle, i.e. the principle of effective elementarity~\cite{elementarity}.

For a single resonance with width, decaying into a single channel, similar principle applies:
the DoS derived from the single channel phase shift contains 
the contribution of the resonance, including its full width, 
and in addition a non-resonant scattering contribution from the asymptotic states~\cite{weinhold,rho,smat,lee}.
The latter tend to dominate at threshold, and give a substantial contribution 
to the thermal pressure due to the smaller Boltzmann suppression.
Within a simple model~\cite{lee,Giacosa:2020tha}, 
one can verify the equivalence between approaches of thermodynamics:
directly computing the partition function with the Hamiltonian via ${\rm Tr} \, e^{-\beta H}$, 
versus the S-matrix formulation.

In this work we start exploring the case of multiple resonances with multiple channels.
In particular, we want to investigate how a dynamically generated state is reflected in the DoS.
To this end, we study a coupled-channel model capable of doing so, 
and study the phase of ${\rm det} \, S$ in the complex plane. 
The DoS is given by the energy derivative along the real line.
We find that the S-matrix DoS is only influenced by those states which are directly connected 
to the physical sheet, 
suggesting not all states are counted in the partition sum.
The influence from other dynamical features, e.g. roots and Riemann sheet structures, will also be examined.

\section{Density of States of a Coupled-Channel System}

\subsection{S-matrix formulation of statistical mechanics}

The S-matrix theory provides a natural language to describe resonances and multi-channel dynamics. 
For the simple case of a single-channel, 2-body interactions, the scattering phase shift $\delta(E)$
uniquely identifies the DoS due to the presence of interactions, via an effective spectral function 

\begin{align}
  B(E) = 2 \frac{\partial}{\partial E} \, \delta(E). 
\end{align}
This effective spectral function defines the thermodynamics of an interacting system\cite{dmb}. Schematically it reads

\begin{align}
  \Delta \ln Z = \int \, \frac{dE}{2 \pi} \, B(E) \, e^{-\beta E}.
\end{align}
Note that the effective spectral function $B(E)$ describes the DoS for $\ln Z$ rather than $Z$. 
This excludes the disconnected diagrams~\cite{dmb}.
As energy increases, new interaction channel opens up and the scattering becomes inelastic. 
The S-matrix should be formally understood as a matrix acting in the space of open channels.
The effective spectral function $B(E)$, in the case of a coupled-channel system, generalizes to~\cite{smat,chibs}

\begin{align}
	\label{eq:B}
		B(E) &= \frac{1}{2} \, {\rm Im} \, {\rm Tr} \, \left[ \, S^{-1} \frac{\partial}{\partial E}  S - \left(\frac{\partial}{\partial E} S^{-1} \right) S \, \right] \\
	             &= \frac{\partial}{\partial E}  \, {\rm Im} \, \ln {\rm det} \, S(E).
\end{align}
where $S$ is an $N_{ch} \times N_{ch}$ S-matrix of the coupled-channel system.

The quantity $B(E)$ summarizes the interactions among the scattering channels. 
For example, inelastic processes, expressed by the off-diagonal S-matrix elements, are included via the determinant.
This poses strong theoretical constraints in model studies: 
when an inelastic process $ \alpha \rightarrow \beta $ is considered, it is necessary to consider also the processes:
$ \beta \rightarrow \alpha $ and $ \beta \rightarrow \beta $, on top of the elastic channel $ \alpha \rightarrow \alpha$.
Nevertheless, the trace operation implies that this quantity is basis independent, i.e., 
two S-matrices related by unitary rotations will give the same DoS.
This also suggests that $B(E)$ does not depend explicitly on the inelasticity parameters.

Based on $B(E)$, an effective phase shift $\mathcal{Q}$ can be constructed:

\begin{align}
\label{eq:Q}
	\begin{split}
\mathcal{Q}(E) &= \frac{1}{2} \, \int_{E_{\rm ref}}^{E} \, d E^\prime \, B(E) \\
	       &= \frac{1}{2} \, {\rm Im} \, \ln {\rm det} \, \left( S(E) / S(E_{\rm ref}) \right). 
	\end{split}
\end{align}
This allows the discussion of a single phase shift function for the whole multi-channel system. 
Note that this quantity is also equivalent to the sum of eigenphases~\cite{weidenmueller}. 

Determining the full N-body, multi-channel S-matrix of an interacting system is in general very difficult (if not impossible). 
Rigorous theoretical schemes, such as chiral perturbation theory~\cite{chipt,Oller:2007aa} and various functional methods, are effective in describing the single-channel, low-energy limit. 
Effective models are adequate for channels dominated by a single, nearby resonance. 
Beyond these cases, inferring the DoS $B(E)$ from individual channels can be rather inefficient. 

\subsection{HRG approximation}

A simple scheme for incorporating resonances in $B(E)$ is the hadron resonance gas (HRG) model~\cite{Hagedorn:1965st,hrgnature}.
Translating into the language of S-matrix, it corresponds to the approximation scheme

\begin{align}
	\label{eq:hrg}
	{\rm det S}(E) = \prod_{\rm \{{\rm res}\}} \, \frac{z_{\rm res}^\star-E}{z_{\rm res}-E},
\end{align}
where $\{\rm res\}$ is a table of resonances (e.g. from PDG) approximated as simple poles

\begin{align}
	z_{\rm res} \approx m_{\rm res} - i \, 0^+.
\end{align}
$\mathcal{Q}_{\rm HRG}$ is then given by a sum of step functions:~\footnote{
	Note that ${\rm Im} \, \ln (m_{\rm res} - E \pm i \, 0^+ ) = \pm \pi \, \theta(E-m_{\rm res})$. 
	It is recommended to use {\it atan2} numerical implementation (available in most programming language, e.g. c++, fortran, and python) to extract the phase, where this relation is automatic.}

\begin{align}
	\label{eq:hrgps}
	\mathcal{Q}(E) \rightarrow \mathcal{Q}_{\rm HRG}(E) = \sum_{\rm res} d_{IJ} \times \pi \, \theta(E-m_{\rm res}),
\end{align}
where $d_{IJ}$ is the degeneracy factor, and the DoS in this case is given by the spectral function $A_{\rm HRG}$:
\begin{align}
	\label{eq:hrgps2}
		B(E) \rightarrow A_{\rm HRG}(E) = \sum_{\rm res} d_{IJ} \times 2 \, \pi \, \delta(E-m_{\rm res}).
\end{align}

One key aspect to improving the approach is by including the widths of the resonances. 
Indeed, considerable theoretical efforts are involved in locating and characterizing resonance poles.
Poles may be identified by analyzing the magnitude of ${\rm det} \, S(E)$. 
Nevertheless, Eq.~(\ref{eq:Q}) urges us to look at the \textbf{phase} of the ${\rm det} \, S(E)$. 
In the following, we shall study the phase function of a familiar 2 channel scattering problem between $\pi\pi, K \bar{K}$. 
By directly graphing this function in the complex plane, we can read off the effective phase shift $\mathcal{Q}(E)$, 
and gain a robust comprehension of how the various S-matrix features: nearby poles, roots, cuts and effects of Riemann surfaces 
appear in the DoS.

\section{DoS in a $\pi\pi, K \bar{K}$ coupled-channel system}

In this section we study the DoS of a $\pi\pi, K \bar{K}$ coupled-channel system. 
The goal is not so much about explaining the scattering data, 
as it is relatively well understood by theoretical approaches, e.g. Refs.~\cite{COLANGELO2001125,PhysRevD.83.074004}.
Instead, we shall make use of this familiar example to investigate 
how the various dynamical structures in the complex plane exert their influences on the DoS.
As we shall see, not all resonances extracted can influence this quantity.

We consider the coupled-channel model in Refs.~\cite{Markushin:2000fa,Locher:1997gr,Morgan:1993td}.
In this approach, an effective Hamiltonian (represented as a $3 \times 3$ matrix) is constructed to describe the interactions 
among the open channels: $\pi\pi, K \bar{K}$, labeled by the channel index $\alpha = 1, 2$, respectively, 
and their coupling to a resonance ($\alpha = 3$).
From this the S-matrix ($2 \times 2$) is derived, which acts only on the open channels.
This type of model is thus particularly suited for investigating how resonances 
and other S-matrix structures arise from the underlying Hamiltonian.

\subsection{constructing the S-matrix}

We work in the center of mass frame and the total energy $E$ is simply given by the invariant mass $\sqrt{s}$.
Our starting point is the free Green's function, which takes the form~\cite{Markushin:2000fa}: 

\begin{align}
	\label{eq:G0_1}
	G^0(s) = {\rm diag} \left[ G^0_{\pi \pi}, \, G^0_{K \bar{K}}, \, \frac{1}{s-M_R^2+i \, 0^+} \right]
\end{align}
where
\begin{align}
	\label{eq:G0_2}
	\begin{split}
		G^0_{\alpha = 1, 2} &= 16 \pi \int \frac{d^3 q^\prime}{(2 \pi)^3} \, \frac{1}{s - (2 \, \epsilon_\alpha)^2 + i \, 0^+} \, R_\alpha(q^\prime),  \\
		\epsilon_1 &= \sqrt{{q^\prime}^2 + m_\pi^2} \\
		\epsilon_2 &= \sqrt{{q^\prime}^2 + m_K^2}.
	\end{split}
\end{align}
A form factor $R_\alpha(q^\prime)$ is needed to render the real part of the integral $G^0_\alpha$ finite. A good choice is

\begin{align}
	R_\alpha(q^\prime) = \left( \frac{\Lambda_\alpha^2}{ \Lambda_\alpha^2 + {q^\prime}^2} \right)^2,
\end{align}
and the integral can be computed analytically:

\begin{align}
	\label{eq:reg}
	\begin{split}
	G^0_\alpha(s) &= \frac{\Lambda_\alpha^3}{2 \, (q_\alpha + i \, \Lambda_\alpha)^2} \\
		    q_{\alpha=1,2}(s)  &= \sqrt{s/4 - m_{\pi, K}^2}.
	\end{split}
\end{align}
Check that as $\Lambda \rightarrow \infty$, we get

\begin{align}
	\begin{split}
		{\rm Im} \, G^0_\alpha(s) &= -q_\alpha + \mathcal{O}(1/\Lambda_\alpha^2) \\
		{\rm Re} \, G^0_\alpha(s) &= -\Lambda_\alpha/2 + \mathcal{O}(1/\Lambda_\alpha).
	\end{split}
\end{align}
Indeed the imaginary part of the integral is finite even without regulation, giving a phase space factor $q_\alpha$. 
For our purpose it is sufficient to treat $\Lambda_\alpha$'s as model parameters, instead of trying to remove them via a subtraction scheme.
This simple prescription also preserves the dispersion relation~\cite{Donoghue:1996kw,Zwicky2016} connecting the real and the imaginary part of the function, i.e. 

\begin{align}
	G^0_\alpha(q^2) &= -\frac{1}{\pi} \, \int_0^\infty d {q^\prime}^2 \, \frac{ {\rm Im} \, G^0_\alpha({q^\prime}^2)}{q^2 - {q^\prime}^2 + i \, 0^+}.
\end{align}

The potential $V$ describes the interactions among the open channels ($\alpha = 1, 2$) and their coupling to the resonance ($\alpha = 3$):

\begin{align}
	V = \left[ \begin{matrix} V_{11} & V_{12} & V_{13} \\ V_{21} & V_{22} & V_{23} \\ V_{31} & V_{32} & V_{33} \end{matrix} \right].
\end{align}
The couplings are in general $s$-dependent. 
Here we follow the parametrization of Ref.~\cite{Markushin:2000fa} (Fit 4). 
For the reader's convenience, we reproduce them here:

\begin{align}
	\begin{split}
		V_{11} &= 5.299 - 2.954 \times s \\
		V_{22} &= - 3.725 \times s \\
		V_{33} &= 0 \\
		V_{12} = V_{21} &= 0.341 \times s \\
		V_{13} = V_{31} &= 2.588 \\
		V_{23} = V_{32} &= 0.702 \\
		\Lambda_1, \Lambda_2 &= 0.529, 0.7 \\
		m_R &= 1.105 \\
		m_\pi, m_K &= 0.1396, 0.4937.
	\end{split}
\end{align}
All parameters are in appropriate units of GeV's. 

The Lippmann-Schwinger equation can be easily solved by matrix inversion: 

\begin{align}
	\begin{split}
		G &= G^0 + G^0 \, V \, G, \\
		T &= V + V \, G^0 \, T.
	\end{split}
\end{align}
A key step is in extracting the S-matrix: This can be achieved by constructing the operator~\cite{smat,lee}:

\begin{align}
	\label{eq:smat}
	\begin{split}
	\tilde{S} &= (I - G^0_- \, V) \, (I + G^0_+ \, T) \\
		&= I - G^0_- \, V + G^0_+ \, T - G^0_- \, V \, G^0_+ \, T \\
		&= I - G^0_- \, V + G^0_+ \, V +  G^0_+ \, V \, G^0_+ \, T - G^0_- \, V \, G^0_+ \, T \\
		&= I + (G^0_+ - G^0_-) \, V  +  (G^0_+ - G^0_-) \, V \, G^0_+ \, T \\
		&= I + (G^0_+ - G^0_-) \, T \\
		&\rightarrow I + 2 \, i \,  {\rm Im} \, (G^0_+) \times T. 
	\end{split}
\end{align}
The last line works for on-shell limit: $E$ is real.~\footnote{ Here we need to follow the convention of Ref.~\cite{Markushin:2000fa} to use their fitting parameters. 
Checking for the diagonal element and note that ${\rm Im} \, G^0_\alpha(s) \sim -q_\alpha $, we see that $ S_{\alpha \alpha} \sim 1 - 2 \, i \, T_{\alpha \alpha}$, identifying 
$T_{\alpha \alpha} = - f_\alpha$ to the scattering amplitude, instead of the standard $T_{\alpha \alpha} = -8 \, \pi \, \sqrt{s} \, f_\alpha$. }
The actual S-matrix $S$ can obtained by projecting the upper $2 \times 2$ subspace of $\tilde{S}$.

The effective spectral function $B(\sqrt{s})$ for the coupled-channel system can be computed based on the determinant of such an S-matrix, 
evaluated on the real line, see Eq.~(\ref{eq:B}). 
As we shall see, the contribution from the resonance (dressed) and other dynamically generated states are naturally 
included / excluded in the physical quantity $B(E)$, or equivalently the phase shift $\mathcal{Q}$.

\begin{figure*}
\includegraphics[width=0.494\linewidth]{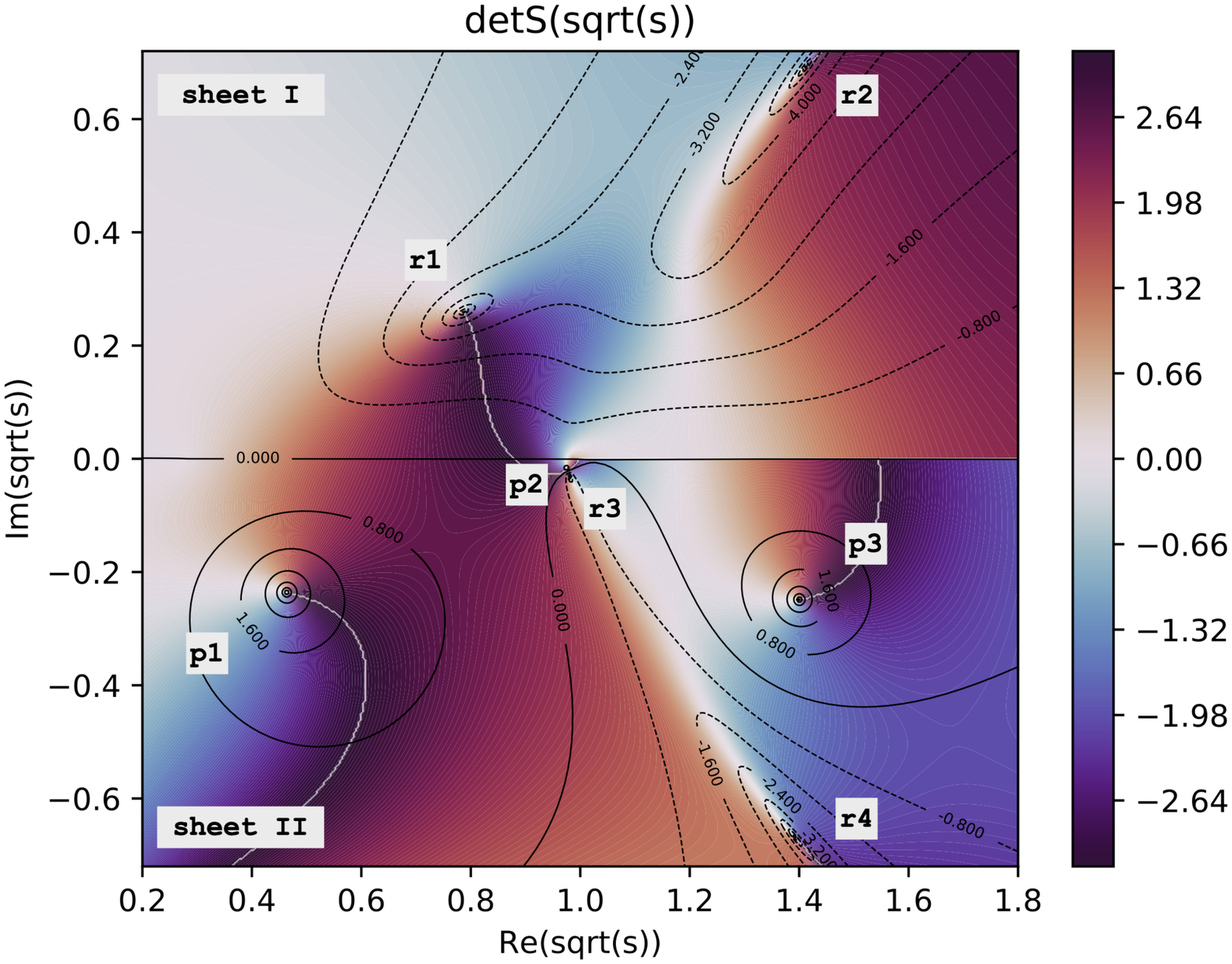}
\includegraphics[width=0.494\linewidth]{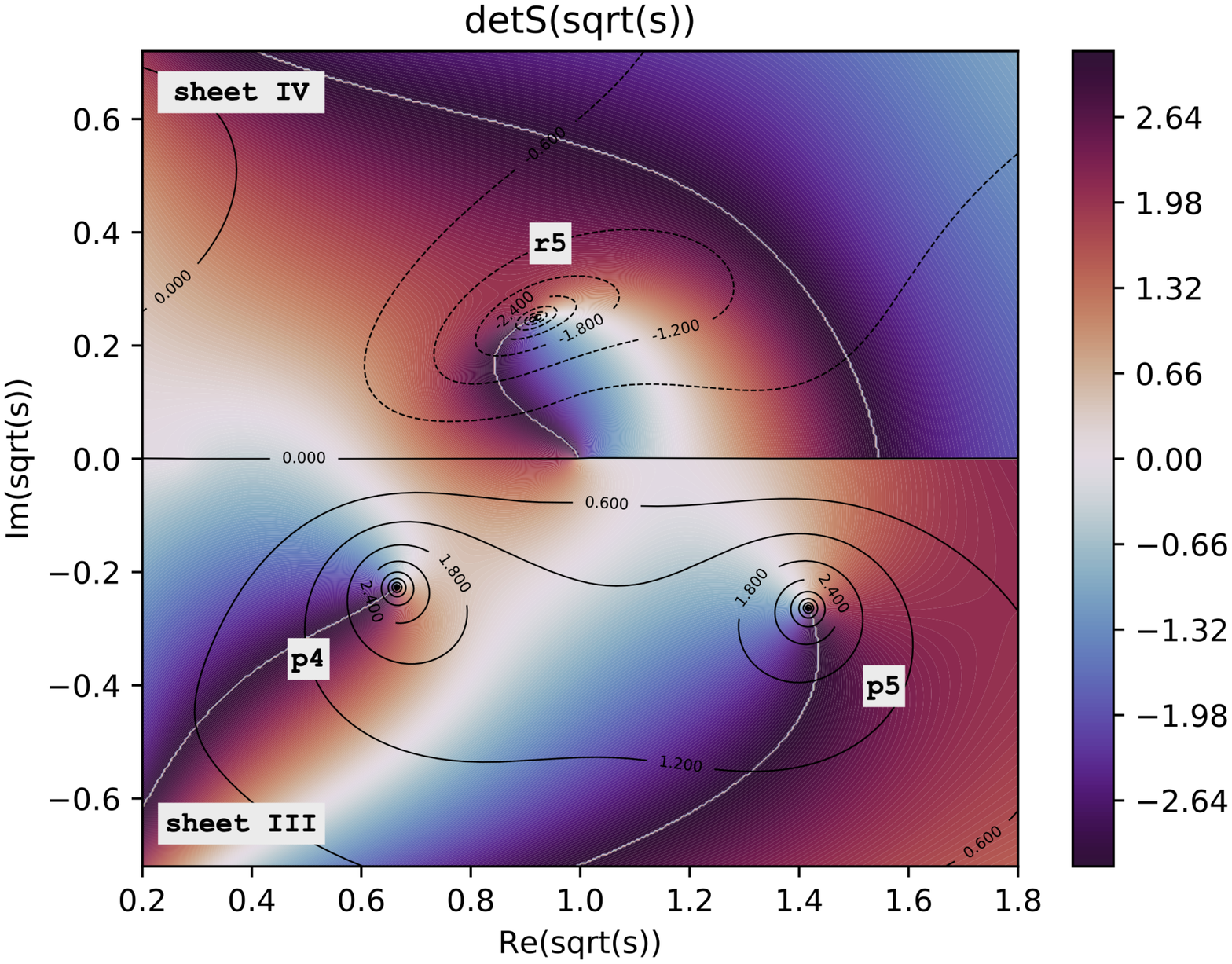}
\caption{Landscape of the phase of the determinant of the S-matrix for the $\pi\pi, K \bar{K}$ coupled-channel system on the energy sheets (left) I (upper) and II (lower) and (right) IV (upper) and III (lower). See Table~\ref{tab1} for the definition of Riemann sheets. 
	Color signifies the value of the phase angle and contour lines specify magnitudes of $ \ln \, \vert  {\rm det} \, S \vert$. 
	Poles (roots) are characterized by the clockwise (anti-clockwise) rotation of the color phase 
	and by a large, positive (negative) values of $ \ln \, \vert  {\rm det} \, S \vert$ reflected in the contour lines.
	The physical line is identified with the real line in sheet I (${\rm Re} \,(\sqrt{s}) + i \, 0^+$). 
	This is where the value of $\mathcal{Q}$ in Eq.~(\ref{eq:Q}) is evaluated. How rapid the phase motion 
	on the physical line determines the magnitude of the density of states. 
	The smoothness of color in transiting the real line indicates the connectedness of the Riemann sheets (see text): 
	between sheet I and II (I and III) below (above) the $K \bar{K}$ threshold. 
	Five resonances can be identified in this model: 3 on sheet II (left, lower half) and 2 on sheet III (right, lower half).
	\label{fig1}}
\end{figure*}

\begin{figure*}
	\includegraphics[width=0.494\linewidth]{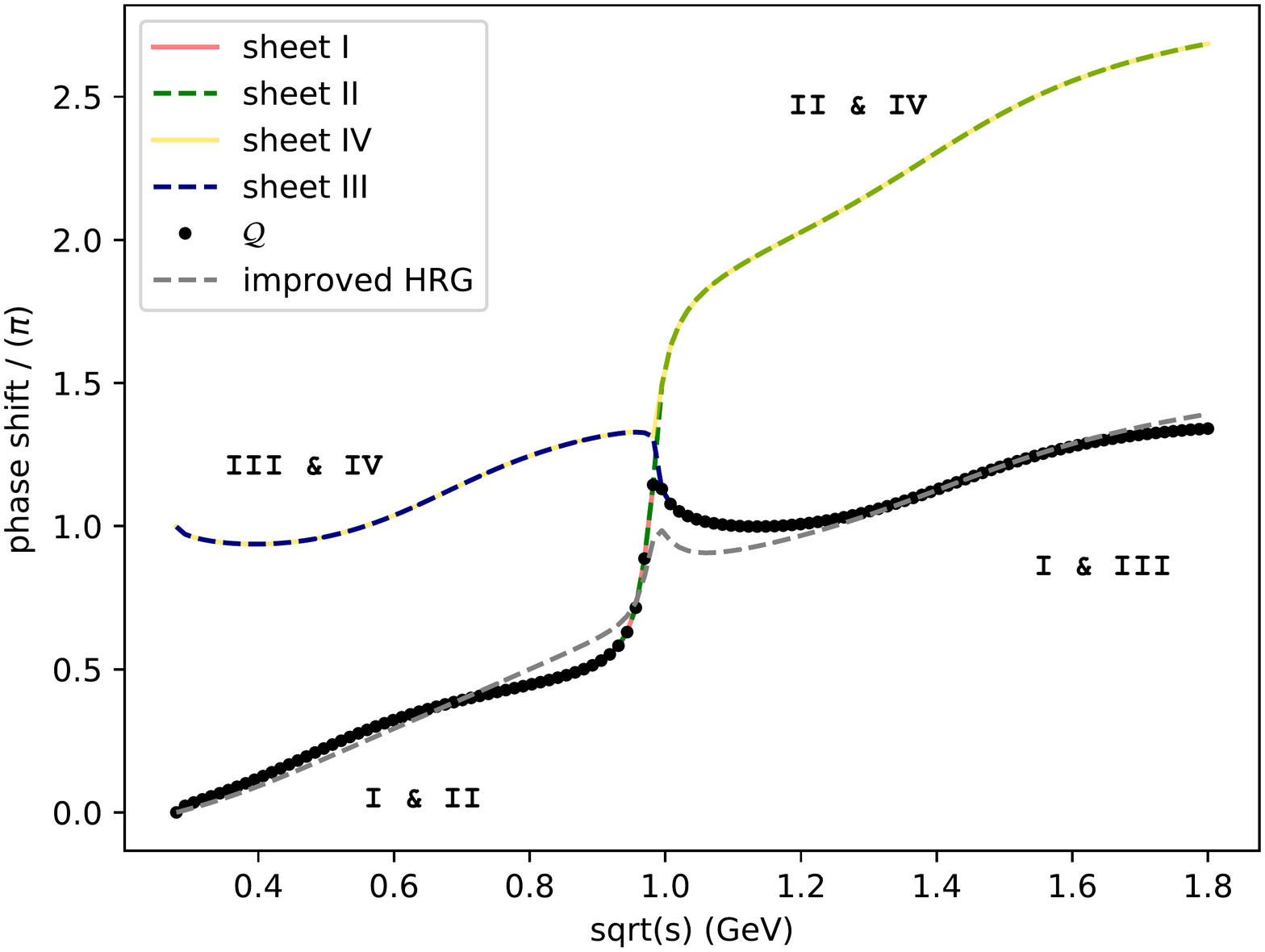}
	\includegraphics[width=0.494\linewidth]{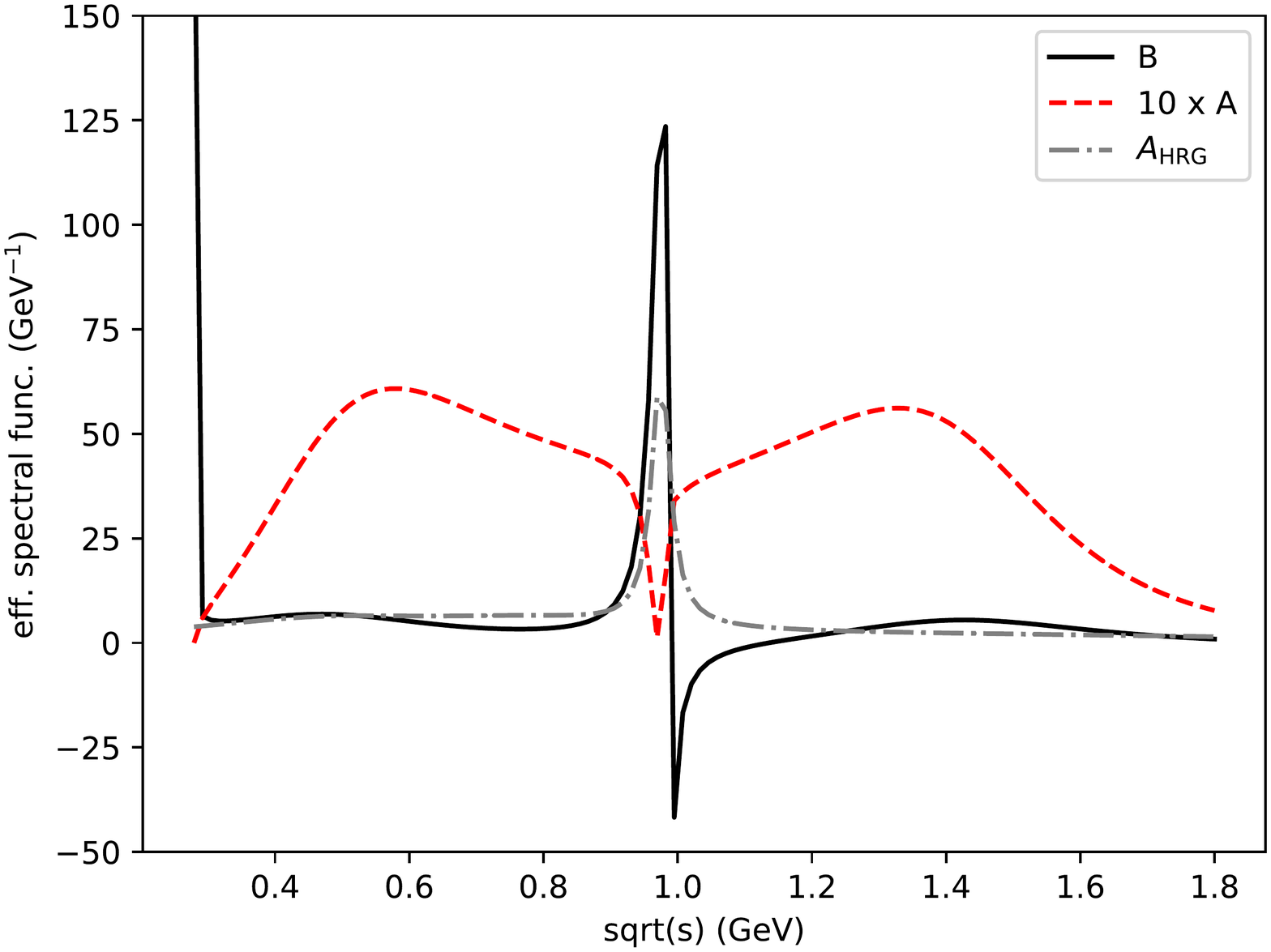}
	\caption{Left: The effective phase shifts for the coupled-channel system. $\mathcal{Q}_B$ is computed using Eq.~(\ref{eq:Q}). 
	It agrees with that extracted from Fig.~\ref{fig1} along sheet I. 
	It also agrees with that extracted from sheet II (III) below (above) the $K \bar{K}$ threshold, 
	as expected from the analysis of the Riemann sheet structure. 
	Note that an extra $2 \pi$ is added to the phase of ${\rm det} \, S$ in sheet III and IV for continuity.
	Also included is the result based on an HRG-like approximation scheme by including only the relevant poles and roots. (see text)
	Right: The effective spectral function $B$ and the spectral functions $A$ of the model. 
	}
\label{fig2}
\end{figure*}

\subsection{phase of ${\rm det} \, S$ in the complex plane}

We compute ${\rm det} \, S$ in the complex plane of $ \sqrt{s} $. 
The 2 branches of $q_{\alpha = 1,2}$ from taking the square root, 
distinguished by the sign of their imaginary parts, 
constitute the 4 energy sheets labeled by (${\rm Im} \, q_1$, ${\rm Im} \, q_2$).
See Table~\ref{tab1}. 

It is natural to expect the values of the complex function ${\rm det}\, S(\sqrt{s})$ in the upper and lower half-planes in a given sheet to be related by complex conjugates, 
as instructed by the Schwarz reflection principle~\cite{Churchill2009} for analytic functions: $F(\bar{z}) = \overline{F(z)}$. 
We numerically verified it is indeed the case for the model. Hence it suffices to show one of the half-planes in each sheet.
In the following, we choose to plot the upper planes of sheet I and IV, and the lower planes of sheet II and III.
As we shall see, this arrangement helps to investigate the connectedness of Riemann sheets.
See appendix~\ref{sec:app} for the motivation behind.

The complex function is shown in Fig.~\ref{fig1}. 
This method of visualizing a complex function is similar to the technique of domain coloring (instead of using brightness we show contours), see Ref.~\cite{wegert2012visual} for details.
The value of the phase angle is represented as color, and the contour lines signify the magnitude of $ \ln \, \vert  {\rm det} \, S \vert$.
This way, poles (roots) are characterized by the clockwise (anti-clockwise) rotation of the color phase 
and by a large, positive (negative) values of the magnitude function. 

Physical observables are extracted along the real line of $\sqrt{s}$.
Thus the actual influence of an S-matrix object on physical quantities depends on its distance to the real line and the Riemann sheet it is on. 
In close proximity a pole (or root) causes rapid phase motion, well described by a standard Breit-Wigner treatment.
When probed afar the influence becomes a non-trivial background, which in some cases can be significant.
This point is obvious in Fig.~\ref{fig1}: for the same horizontal distance ($\Delta E$), the phase (color) change is much more rapid near the source. 
It is also clear that roots act as anti-poles with regard to phase motion. 
A pole and a root lying close to each other will tend to neutralize the phase motion outside the pair -- 
a situation similar to arrangement of electric charges.
In fact, the Cauchy's argument principle~\cite{Churchill2009} relates the change in phase around a loop to 
the difference between the total number of poles versus that of roots. 
\textbf{For phase motion, roots are as important as poles.}

Unlike the K-matrix approach, the coupled-channel model considered here allows the dynamical generation of resonances. 
Starting from a single bare resonance state, a total of 5 resonance poles can be identified in the model. See Table~\ref{tab2}.
In particular $p1, p2$ are the familiar resonances $\sigma(500), f_0(980)$. 
The resonances are distributed across the Riemann sheets II (3 poles) and III (2 poles).
It turns out that the phase value on the physical line (the color at ${\rm Re} \,(\sqrt{s}) + i \, 0^+$) is only strongly affected by 3 out of the 5 poles: $p1$, $p2$, and $p5$. 
This is naturally understood when considering the \textbf{continuity} of the phase of ${\rm det} \, S$ across the Riemann sheets. 

The continuity can be understood as follows: In going across the real line from $i \, 0^+$ to $-i \, 0^+$, 
for ${\rm Re} \, \sqrt{s}$ below the $K \bar{K}$ threshold, we essentially travel from sheet I to sheet II. 
This is indicated by the smooth color change across the real line, i.e. (top left to bottom left, left figure). Similar observation is made from sheet IV to sheet III, i.e. (top left to bottom left, right figure). Above the threshold, however, this becomes a transition from sheet I to sheet III. In this case, color changes smoothly across the real line from sheet I to sheet III, i.e. (top right, left figure) to (bottom right, right figure). For the latter case, objects on sheet II (e.g. $p3$ pole) would barely influence the real energy line. A similar conclusion is made for objects in sheet III (e.g. $p4$ pole) in the former case.
In any case, the smoothness of color in transiting the real line indicates the connectedness of the Riemann sheets.
This gives an intuitive criteria for the relevance of poles (and roots) in the complex plane when calculating the physical DoS.

S-matrix roots are also important in determining the phase motion in the physical sheet. See Table~\ref{tab2}.
The importance of roots on channel amplitudes have been stressed in previous analyses~\cite{Kaminski:1998ns,Kaminski:1993zb}.
For the density of state, we find substantial contribution from the root: $r1, r2$, and $r3$. 
The roots $r4$ in sheet II and $r5$ in sheet IV, with the continuity argument, can be safely neglected.

The phase shift function $\mathcal{Q}$ in Eq.~(\ref{eq:Q}) encapsulates the effects from the various S-matrix objects.
This quantity is evaluated on the physical line, and can be directly read off from Fig.~\ref{fig1}: $(1/2)$ the value of the phase function along the real line of sheet I.
From the Riemann sheet structure, we expect the same result can be extracted from sheet II (III) below (above) the $K \bar{K}$ threshold. 
Last but not least, it can be directly evaluated via an integral of the effective spectral function $B$ along the real line. (first line of Eq.~(\ref{eq:Q}))
The last option seems to be superfluous, but is in fact one of the most useful. 
For one thing, it is free of the ambiguity of adding / subtracting multiples of the $2 \pi$ when computing the phase function. 
Also, the effective spectral function $B$ in Eq.~(\ref{eq:B}) is generally continuous, though sometimes (integrably) diverging at thresholds~\cite{kappa}.
It is reassuring to see the various ways of extracting $\mathcal{Q}$ to agree. See Fig.~\ref{fig2} (Left).

\begin{table}[t]
	\begin{minipage}{0.48 \textwidth}
		\centering
\begin{tabular}{|c|c|c|}
  \hline
	\normalsize {}     &  ${\rm Im} \, q_{\pi \pi} $       &  ${\rm Im} \, q_{K \bar{K}} $ \\ \hline
	sheet I            &  +                                      &   +        			   \\ \hline
	sheet II           &  -                                      &   +                                 \\ \hline
        sheet III          &  -                                      &   -                                 \\ \hline
	sheet IV           &  +                                      &   -                                 \\ \hline
\end{tabular}
	 \caption{Definition of Riemann sheets. Convention follows Ref.~\cite{Frazer:1964zz,riemann_shts}}
	\label{tab1}
	\end{minipage}
\hfill
	\begin{minipage}{0.48\textwidth}
		\centering
\begin{tabular}{|c|c|c|c|}
  \hline
	\normalsize {}      &  ${\rm Re} \, \sqrt{s} $       &  ${\rm Im} \, \sqrt{s} $ & sheet \\ \hline
	p1                  &  0.4637 & -0.2357 & II            \\ \hline
	p2                  &  0.975  & -0.0164 & II            \\ \hline
	p3                  &  1.401  & -0.249  & II            \\ \hline
	p4                  &  0.6654 & -0.2263 & III           \\ \hline
	p5                  &  1.4176 & -0.2640 & III           \\ \hline \hline
	r1		    &  0.787  & +0.259  & I             \\ \hline
	r2		    &  1.410  & +0.691  & I             \\ \hline
	r3		    &  0.981  & -0.032  & II            \\ \hline
	r4		    &  1.393  & -0.669  & II            \\ \hline
	r5		    &  0.918  & +0.248  & IV            \\ \hline
\end{tabular}
	\caption{Location of resonance poles ($p_i$) and roots ($r_i$) identified in the model.}
	\label{tab2}
	\end{minipage}
\end{table}

\subsection{improving the HRG approximation}

We now examine the efficacy of several approximations for the DoS.
The exact result in this model can be easily calculated by:

\begin{align}
	B(\sqrt{s}) = 2 \, \frac{\partial}{\partial \sqrt{s}} \, \mathcal{Q}(\sqrt{s}),
\end{align}
or be obtained from the trace of products of S-matrix via Eq.~(\ref{eq:B}). See Fig.~\ref{fig2} (Right). 
A prominent feature of the effective spectral function $B$ is the divergence at thresholds.
This appears for S-wave scattering, where $\mathcal{Q} \approx a_S \times q(s)$ close to thresholds, and 

\begin{align}
	\label{eq:thres}
	\begin{split}
		B &\approx 2 \, a_S \, \frac{\partial}{\partial \sqrt{s}} \, \sqrt{s/4-m^2_{\pi, K}} \\
		    	    &= 2 \, a_S \, \frac{\sqrt{s}/2}{\sqrt{s/4-m^2_{\pi, K}}}.
	\end{split}
\end{align}
Clearly it diverges when $\sqrt{s} \rightarrow 2 \, m_{\pi, K}$, behaving like $\propto 1/q$. 
Note that the sign of the divergence mirrors that of the scattering length $a_S$.
The first line of Eq.~(\ref{eq:thres}) also makes it clear that $\int d \sqrt{s} \, B(\sqrt{s})$ is finite, i.e. the divergence is integrable.

Other than the threshold effects, 
we observe that the $\sigma(500)$-contribution is strongly suppressed, 
significantly less than even a Breit-Wigner treatment. 
(Further negative contribution comes from the $I=2$ sector.)
This leads to the suggestion that this state should be excluded in the thermal model~\cite{sigma, kappa}.
On the other hand, the phase motion from the $p2$ pole, i.e. the $f_0(980)$ resonance, is clearly visible.

When a single narrow resonance dominates the interaction, it is common to approximate $B$ by the spectral function $A$ of the resonance.
The latter is obtained by:
\begin{align}
	A = (2 \sqrt{s}) \times  (-2 \, {\rm Im} \, G^+_{3 3}). 
\end{align}
In this model the spectral function $A$ of the dressed resonance deviates significantly from the DoS. See Fig.~\ref{fig2} (Right). 
In particular it is not dominated by the $f_0(980)$ resonance: 
This is expected since the interacting system is dominated by the dynamically generated $f_0(980)$, rather than by the original seed resonance.

From Fig.~\ref{fig2} it is obvious that a naive application of the HRG approximation scheme in Eqs.~(\ref{eq:hrgps}) and (\ref{eq:hrgps2}) will not work: 
a sum of step functions will not adequately describe the phase shift function $\mathcal{Q}$.
Fortunately, in this simple model we have complete information of the resonance poles and hence correcting for widths is straightforward. 
However including all the resonances in Table~\ref{tab2} will still significantly overestimate the DoS.

Based on our analysis on the Riemann sheet structure, 
we propose a novel (HRG-like) approximation scheme for describing dynamics: 
by patching together \textbf{relevant} S-matrix objects, e.g. resonances, roots, and possibly branch points and cuts.
A rudimentary example for the current model is given by:

\begin{align}
	\label{eq:hrgselect}
	{\rm det} \, S(E) = \begin{dcases*}
		{\frac{(r1-E)(r2-E)}{(p1-E)(p2-E)}} & \text{for} $E < 2 \, m_K$ \\
		\frac{(r1-E)(r2-E)(r3-E)}{(p5-E)} & \text{for} $E \geq 2 \, m_K$
	\end{dcases*}
\end{align}
The selection of poles and roots is based on their connection to the physical line (sheet I). 
The exception is for $r3$, which is a rather unusual root lying on sheet II. 
It lies very close the $p2$ pole, and gives a strong subtractive contribution to the DoS. 
Unlike the spectral function $A$, negative contributions are allowed in forming $B$, 
as it measures the \textbf{change} in the DoS due to the interactions, i.e. corrections compared to the free (2-body) scattering state.
A prime example for a negative contribution to the DoS is repulsions among hadrons, e.g. an excluded volume effect~\cite{Vovchenko:2017drx,exclvol}.
Here we see an alternative source: due to the dynamical generation of roots. 

The $r3$ pole also lies very close to the $K \bar{K}$ threshold. 
Empirically, we find a much better fit if we assign $r3$ to the second case in Eq.~(\ref{eq:hrgselect}).
~\footnote{We expect this point, and the arrangement of poles and roots in general, to be model dependent,
and can depend on the prescription (Eq. (\ref{eq:smat})) in extracting the complex landscape of S-matrix.
The result on the physical line, on the other hand, is not expected to change.}
The scheme is shown in Fig.~\ref{fig2}. 
It is comparable to the DoS $B$, although the threshold effects are not reproduced.

By varying the couplings in this model, 
we can infer that the $f_0(980)$ resonance starts off being a shallow bound state, 
generated within the $K \bar{K}$ channel. 
It then becomes a resonance due to the coupling to the $\pi \pi$ channel.
The characteristic phase motion on $\mathcal{Q}$, i.e. 
a rapid rise similar to a standalone narrow state, accompanied by a rapid drop soon after, 
appears in many states with molecular origin.
Nevertheless, it should be stressed that these observations are essentially model dependent.
A potential way to make progress is by studying the landscape of the phase of ${\rm det} \, S$ in different models, 
and tracking its changes when model parameters are altered.

Even for this simple system we uncover a rich dynamics in the complex plane.
In particular, the phase function gives an intuitive account of how a cluster of poles and roots modifies the DoS. 
The exploratory nature of this study most likely means that the analytic continuation~\cite{Kaminski:1998ns,Kaminski:2019bep} of the S-matrix into the complex plane needs to be significantly improved.
Also there are more dynamical structures in the S-matrix than studied here:
e.g. coupled-channel cusps~\cite{cusp}, logarithmic divergences induced from triangle diagram~\cite{triangle}, etc.
It is not yet clear how they will enter the DoS and is a subject of future research.

\section{Going Further}
\label{sec5}

Studying the phase of the determinant of the scattering matrix in the complex plane reveals rich particle dynamics: 
poles, roots and Riemann sheet structures, 
and give an intuitive account of how they contribute to the density of states (DoS). 
Of course, models become unreliable in describing high-energy processes, 
when multiple channels open up and $N>2$ scatterings are involved.
The aim here is to build on these insights to inventing a more robust approximation scheme. 
For example, based on the analysis of the Riemann sheets and the continuity of the effective phase shift function, 
we can select the most relevant poles and roots in the complex plane, 
and build an approximation scheme analogous to the hadron resonance gas (HRG) model.

Given the large number of predicted states (e.g. by LQCD~\cite{LQCD}) which are unobserved, 
and the observed states which are unconfirmed in experiments, 
a criterion for selecting the most relevant states in constructing the DoS is urgently needed~\cite{
Bazavov:2014xya,missS}, 
This is also essential for reliably computing the bulk properties of thermal medium at high temperatures and densities~\cite{chibs}.
We can imagine a scenario where some of the bound states predicted from a quark model calculation, 
after coupling to the continuum (i.e., open channels), 
they are so redistributed that they have little influence on the DoS and 
will therefore not be counted in the thermal trace.
A word of caution: This may be erroneously taken as a repulsive correction.
Further work needs to be done to fully understand the change in the DoS (and the thermal sum) 
when unquenching the quark model~\cite{Swanson_2005,Coito:2014dya,Rupp:2015taa}.

Up to now we have been arguing for the relevance of a state based on its proximity to the physical line. 
This is a natural requirement but more work is needed to explore the consequences.
In particular, poles (and other S-matrix objects) 
``move'' when the effective interaction strengths are altered.
The change may be induced by varying the quark mass or unquenching the calculations, in the context of LQCD; 
or brought about from the presence of an external electric / magnetic field, in a condensed matter system.
At the moment we can only study case by case, and no general rule is known for deciding the fate of a state.

It would also be useful to explore other interacting system. 
Of course, more sophisticated coupled-channel models exist to describe a variety of systems.
Computing the DoS would require the complete S-matrix: 
including channels where experimental data are unavailable, 
and the dummy channels for restoring unitarity.
A study exploring the thermodynamics of hyperons, based on the DoS extracted from a coupled-channel PWA, can be found in Ref.~\cite{chibs}.

Studying the thermal properties of the system of exotics\cite{exotica,xyz1,xyz2} 
is also possible using this approach. 
A good example of this is the X(3872) system. 
Previous works have focused on the single channel phase shift~\cite{Ortega:2017hpw,Giacosa:2019zxw}. 
Our study here suggests that analyzing the structure of ${\rm det} \,S$ 
may yield a better understanding for characterizing the density of states.

The S-matrix framework discussed in this work is very flexible: 
the degrees of freedom (DoFs) used in the Hamiltonian 
can be different from those appearing in the S-matrix.
Such a separation would become interesting when quarks and gluons DoFs are employed 
in the Hamiltonian~\cite{PhysRevD.11.257,PhysRev.111.995,Machavariani_2011}, 
while the S-matrix is employed in computing the DoS.
Presumably at low temperatures, probing low energies due to Boltzmann suppression, 
it should yield a gas of pions (or pion concepts~\cite{shankar1997effective,Chew:1962mpd}).
Working out the details in the S-matrix approach could yield novel insights into 
describing the thermal properties of interacting hadrons 
and eventually the deconfinement phase transition in QCD~\cite{Bazavov:2011nk}.

\begin{acknowledgments}
The author thanks Eric Swanson for carefully reading the manuscript and for giving constructive comments.
He is also grateful to Valeri Markushin for the kind correspondence, and for sharing ideas on possible future works.
The discussion with Thomas Kl{\"a}hn on fractals inspires this work.
He also benefits from the discussions with (and lectures from) Hans Feldmeier and Bengt Friman at GSI, 
and the productive collaboration in Wroc{\l}aw with Chihiro Sasaki and Krzysztof Redlich.
This study receives supports from the Short Term Scientific Mission (STSM) program under COST Action CA15213 (reference number: 41977) and the Polish National Science Center (NCN) under the Opus grant no. 2018/31/B/ST2/01663.
\end{acknowledgments}

\appendix*
\label{sec:app}
\section{stitching together Riemann Sheets}

\begin{figure*}
	\includegraphics[width=0.9\linewidth]{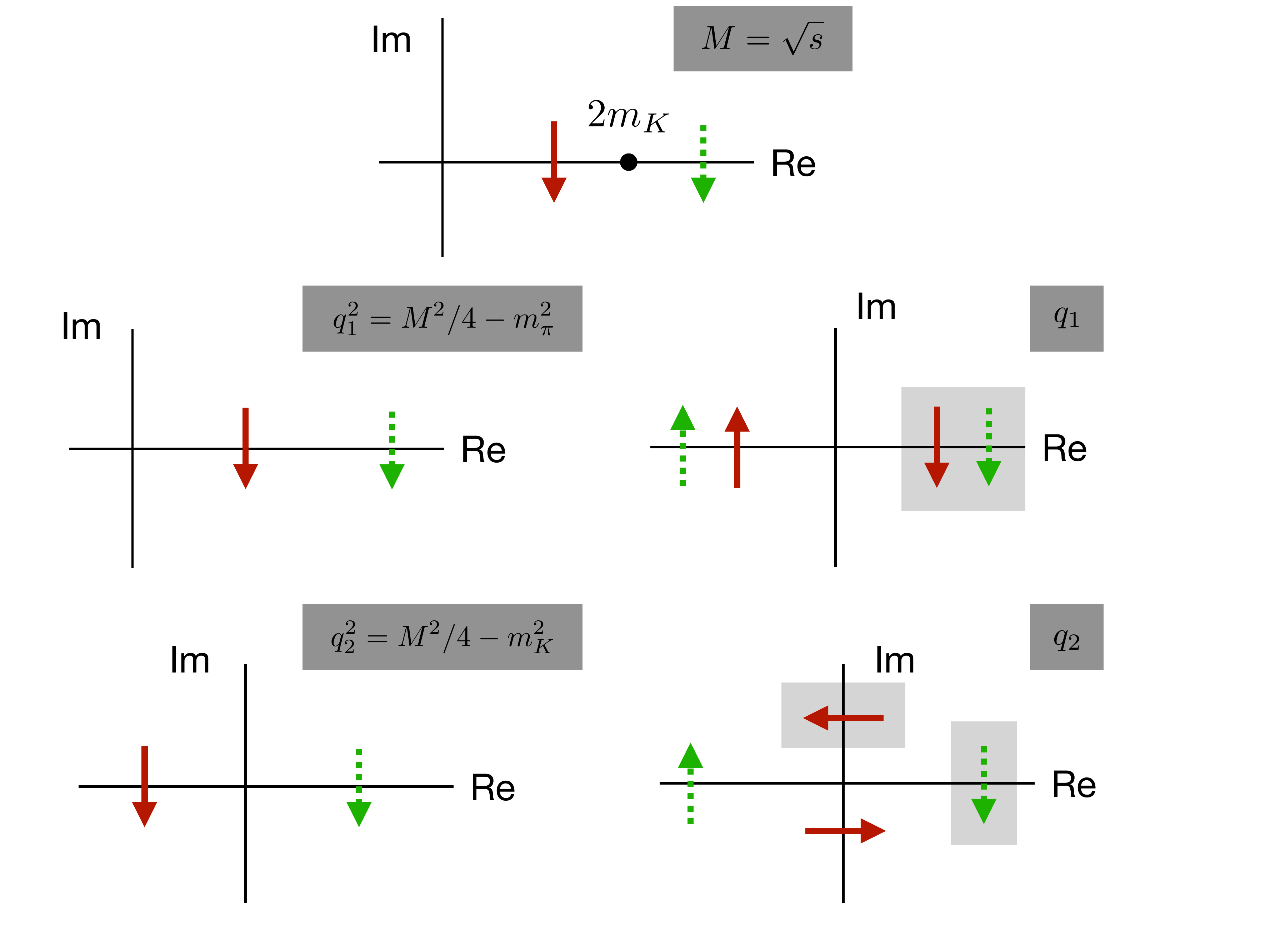}
	\caption{The motions of various dynamical variables as $\sqrt{s}$ moves across the real line in the complex plane.
	\label{fig3}}
\end{figure*}

We shall motivate an arrangement of half-sheets that aids the investigation of the 
continuity of the phase of ${\rm det} \, S$ across the real line. 
The selection is: the upper planes of sheet I and IV, and the lower planes of sheet II and III. 

First it is natural to explore the upper plane of sheet I, as it is where the physical line belongs.
Consider going across the real line below (above) the 2-kaon threshold in the manner depicted in Fig.~\ref{fig3} (top), shown in red (green) arrow.
The corresponding motions of the square of the channel momenta $q_{\alpha = 1,2}$ are also displayed.
The two branches of $q_{\alpha = 1,2}$ are related to each other by multiplying $-1$.
In the complex plane, this corresponds to a reflection about the origin.

In the calculation we always take the positive root of $q_1 = \sqrt{(x+i y)^2/4-m_\pi^2}$. (shaded in grey)
This way, the sign of ${\rm Im} \, q_1$ essentially follows that of $y$. 
For the motion below the 2-kaon threshold (red arrow), we expect only the imaginary part of $q_1$ to flip sign, 
while that of $q_2$ stays: e.g., going from upper (+,+) (sheet I) to lower (-,+) (sheet II). 
Above the 2-kaon threshold (green arrow), the imaginary parts of both $q_\alpha$'s should flip sign: 
e.g., going from upper (+,+) (sheet I) to lower (-,-) (sheet III).

We then perform 2 calculations in the complex plane: 
choosing $q_2 = \pm \sqrt{(x+i y)^2/4-m_K^2}$ for ${\rm Im} \, q_2 > 0$ in one case and ${\rm Im} \, q_2 < 0$ in another.
This binds sheets (I, upper and II, lower) and sheets (IV, upper and III, lower) together, 
and explains the sheet arrangement in Fig.~\ref{fig1}.
Thus the transition of phase (color) across the real line is smooth below 2-kaon threshold.
Above the threshold, one needs to exchange the lower half of the two graphs to see the continuity in the phase in 
sheets (I and III) and sheets (IV and II).
In any case, the smoothness of color in 
transiting the real line indicates the connectedness of the Riemann sheets.

Note that in common numerical implementation (e.g. c++, fortran, python) the positive square root, i.e. positive real part, of a complex number is returned.
In this case $q_1$ can be computed directly but ${\rm Im} \, q_2 > 0$ and ${\rm Im} \, q_2 < 0$ need to be separately implemented.

\bibliography{ref}

\begin{thebibliography}{78}%
\makeatletter
\providecommand \@ifxundefined [1]{%
 \@ifx{#1\undefined}
}%
\providecommand \@ifnum [1]{%
 \ifnum #1\expandafter \@firstoftwo
 \else \expandafter \@secondoftwo
 \fi
}%
\providecommand \@ifx [1]{%
 \ifx #1\expandafter \@firstoftwo
 \else \expandafter \@secondoftwo
 \fi
}%
\providecommand \natexlab [1]{#1}%
\providecommand \enquote  [1]{``#1''}%
\providecommand \bibnamefont  [1]{#1}%
\providecommand \bibfnamefont [1]{#1}%
\providecommand \citenamefont [1]{#1}%
\providecommand \href@noop [0]{\@secondoftwo}%
\providecommand \href [0]{\begingroup \@sanitize@url \@href}%
\providecommand \@href[1]{\@@startlink{#1}\@@href}%
\providecommand \@@href[1]{\endgroup#1\@@endlink}%
\providecommand \@sanitize@url [0]{\catcode `\\12\catcode `\$12\catcode
  `\&12\catcode `\#12\catcode `\^12\catcode `\_12\catcode `\%12\relax}%
\providecommand \@@startlink[1]{}%
\providecommand \@@endlink[0]{}%
\providecommand \url  [0]{\begingroup\@sanitize@url \@url }%
\providecommand \@url [1]{\endgroup\@href {#1}{\urlprefix }}%
\providecommand \urlprefix  [0]{URL }%
\providecommand \Eprint [0]{\href }%
\providecommand \doibase [0]{http://dx.doi.org/}%
\providecommand \selectlanguage [0]{\@gobble}%
\providecommand \bibinfo  [0]{\@secondoftwo}%
\providecommand \bibfield  [0]{\@secondoftwo}%
\providecommand \translation [1]{[#1]}%
\providecommand \BibitemOpen [0]{}%
\providecommand \bibitemStop [0]{}%
\providecommand \bibitemNoStop [0]{.\EOS\space}%
\providecommand \EOS [0]{\spacefactor3000\relax}%
\providecommand \BibitemShut  [1]{\csname bibitem#1\endcsname}%
\let\auto@bib@innerbib\@empty
\bibitem [{\citenamefont {Dashen}\ \emph {et~al.}(1969)\citenamefont {Dashen},
  \citenamefont {Ma},\ and\ \citenamefont {Bernstein}}]{dmb}%
  \BibitemOpen
  \bibfield  {author} {\bibinfo {author} {\bibfnamefont {R.}~\bibnamefont
  {Dashen}}, \bibinfo {author} {\bibfnamefont {S.-K.}\ \bibnamefont {Ma}}, \
  and\ \bibinfo {author} {\bibfnamefont {H.~J.}\ \bibnamefont {Bernstein}},\
  }\href {\doibase 10.1103/PhysRev.187.345} {\bibfield  {journal} {\bibinfo
  {journal} {Phys. Rev.}\ }\textbf {\bibinfo {volume} {187}},\ \bibinfo {pages}
  {345} (\bibinfo {year} {1969})}\BibitemShut {NoStop}%
\bibitem [{\citenamefont {Venugopalan}\ and\ \citenamefont
  {Prakash}(1992)}]{prakash}%
  \BibitemOpen
  \bibfield  {author} {\bibinfo {author} {\bibfnamefont {R.}~\bibnamefont
  {Venugopalan}}\ and\ \bibinfo {author} {\bibfnamefont {M.}~\bibnamefont
  {Prakash}},\ }\href {\doibase https://doi.org/10.1016/0375-9474(92)90005-5}
  {\bibfield  {journal} {\bibinfo  {journal} {Nuclear Physics A}\ }\textbf
  {\bibinfo {volume} {546}},\ \bibinfo {pages} {718 } (\bibinfo {year}
  {1992})}\BibitemShut {NoStop}%
\bibitem [{\citenamefont {LeClair}(2007)}]{leclair1}%
  \BibitemOpen
  \bibfield  {author} {\bibinfo {author} {\bibfnamefont {A.}~\bibnamefont
  {LeClair}},\ }\href {\doibase 10.1088/1751-8113/40/31/033} {\bibfield
  {journal} {\bibinfo  {journal} {Journal of Physics A: Mathematical and
  Theoretical}\ }\textbf {\bibinfo {volume} {40}},\ \bibinfo {pages} {9655}
  (\bibinfo {year} {2007})}\BibitemShut {NoStop}%
\bibitem [{\citenamefont {Lo}(2017)}]{smat}%
  \BibitemOpen
  \bibfield  {author} {\bibinfo {author} {\bibfnamefont {P.~M.}\ \bibnamefont
  {Lo}},\ }\href {\doibase 10.1140/epjc/s10052-017-5106-0} {\bibfield
  {journal} {\bibinfo  {journal} {Eur. Phys. J. C}\ }\textbf {\bibinfo {volume}
  {77}},\ \bibinfo {pages} {533} (\bibinfo {year} {2017})},\ \Eprint
  {http://arxiv.org/abs/1707.04490} {arXiv:1707.04490 [hep-ph]} \BibitemShut
  {NoStop}%
\bibitem [{\citenamefont {Tanabashi}\ \emph {et~al.}(2018)\citenamefont
  {Tanabashi} \emph {et~al.}}]{pdg}%
  \BibitemOpen
  \bibfield  {author} {\bibinfo {author} {\bibfnamefont {M.}~\bibnamefont
  {Tanabashi}} \emph {et~al.} (\bibinfo {collaboration} {Particle Data
  Group}),\ }\href {\doibase 10.1103/PhysRevD.98.030001} {\bibfield  {journal}
  {\bibinfo  {journal} {Phys. Rev. D}\ }\textbf {\bibinfo {volume} {98}},\
  \bibinfo {pages} {030001} (\bibinfo {year} {2018})}\BibitemShut {NoStop}%
\bibitem [{\citenamefont {Gasser}\ and\ \citenamefont
  {Leutwyler}(1984)}]{Gasser:1983yg}%
  \BibitemOpen
  \bibfield  {author} {\bibinfo {author} {\bibfnamefont {J.}~\bibnamefont
  {Gasser}}\ and\ \bibinfo {author} {\bibfnamefont {H.}~\bibnamefont
  {Leutwyler}},\ }\href {\doibase 10.1016/0003-4916(84)90242-2} {\bibfield
  {journal} {\bibinfo  {journal} {Annals Phys.}\ }\textbf {\bibinfo {volume}
  {158}},\ \bibinfo {pages} {142} (\bibinfo {year} {1984})}\BibitemShut
  {NoStop}%
\bibitem [{\citenamefont {Oller}\ and\ \citenamefont {Oset}(1997)}]{Oller}%
  \BibitemOpen
  \bibfield  {author} {\bibinfo {author} {\bibfnamefont {J.~A.}\ \bibnamefont
  {Oller}}\ and\ \bibinfo {author} {\bibfnamefont {E.}~\bibnamefont {Oset}},\
  }\href {\doibase 10.1016/S0375-9474(99)00427-3,
  10.1016/S0375-9474(97)00160-7} {\bibfield  {journal} {\bibinfo  {journal}
  {Nucl. Phys.}\ }\textbf {\bibinfo {volume} {A620}},\ \bibinfo {pages} {438}
  (\bibinfo {year} {1997})},\ \bibinfo {note} {[Erratum: Nucl.
  Phys.A652,407(1999)]},\ \Eprint {http://arxiv.org/abs/hep-ph/9702314}
  {arXiv:hep-ph/9702314 [hep-ph]} \BibitemShut {NoStop}%
\bibitem [{\citenamefont {Shepherd}\ \emph {et~al.}(2016)\citenamefont
  {Shepherd}, \citenamefont {Dudek},\ and\ \citenamefont
  {Mitchell}}]{Shepherd:2016dni}%
  \BibitemOpen
  \bibfield  {author} {\bibinfo {author} {\bibfnamefont {M.~R.}\ \bibnamefont
  {Shepherd}}, \bibinfo {author} {\bibfnamefont {J.~J.}\ \bibnamefont {Dudek}},
  \ and\ \bibinfo {author} {\bibfnamefont {R.~E.}\ \bibnamefont {Mitchell}},\
  }\href {\doibase 10.1038/nature18011} {\bibfield  {journal} {\bibinfo
  {journal} {Nature}\ }\textbf {\bibinfo {volume} {534}},\ \bibinfo {pages}
  {487} (\bibinfo {year} {2016})},\ \Eprint {http://arxiv.org/abs/1802.08131}
  {arXiv:1802.08131 [hep-ph]} \BibitemShut {NoStop}%
\bibitem [{\citenamefont {Rapp}\ and\ \citenamefont
  {Wambach}(2000)}]{Rapp:1999ej}%
  \BibitemOpen
  \bibfield  {author} {\bibinfo {author} {\bibfnamefont {R.}~\bibnamefont
  {Rapp}}\ and\ \bibinfo {author} {\bibfnamefont {J.}~\bibnamefont {Wambach}},\
  }\href {\doibase 10.1007/0-306-47101-9_1} {\bibfield  {journal} {\bibinfo
  {journal} {Adv. Nucl. Phys.}\ }\textbf {\bibinfo {volume} {25}},\ \bibinfo
  {pages} {1} (\bibinfo {year} {2000})},\ \Eprint
  {http://arxiv.org/abs/hep-ph/9909229} {arXiv:hep-ph/9909229 [hep-ph]}
  \BibitemShut {NoStop}%
\bibitem [{\citenamefont {Godfrey}\ and\ \citenamefont
  {Isgur}(1985)}]{Godfrey:1985xj}%
  \BibitemOpen
  \bibfield  {author} {\bibinfo {author} {\bibfnamefont {S.}~\bibnamefont
  {Godfrey}}\ and\ \bibinfo {author} {\bibfnamefont {N.}~\bibnamefont
  {Isgur}},\ }\href {\doibase 10.1103/PhysRevD.32.189} {\bibfield  {journal}
  {\bibinfo  {journal} {Phys. Rev. D}\ }\textbf {\bibinfo {volume} {32}},\
  \bibinfo {pages} {189} (\bibinfo {year} {1985})}\BibitemShut {NoStop}%
\bibitem [{\citenamefont {Barnes}\ and\ \citenamefont
  {Swanson}(1992)}]{Barnes1992}%
  \BibitemOpen
  \bibfield  {author} {\bibinfo {author} {\bibfnamefont {T.}~\bibnamefont
  {Barnes}}\ and\ \bibinfo {author} {\bibfnamefont {E.~S.}\ \bibnamefont
  {Swanson}},\ }\href {\doibase 10.1103/PhysRevD.46.131} {\bibfield  {journal}
  {\bibinfo  {journal} {Physical Review D}\ }\textbf {\bibinfo {volume} {46}},\
  \bibinfo {pages} {131} (\bibinfo {year} {1992})}\BibitemShut {NoStop}%
\bibitem [{\citenamefont {Broniowski}\ \emph {et~al.}(2003)\citenamefont
  {Broniowski}, \citenamefont {Florkowski},\ and\ \citenamefont
  {Hiller}}]{Broniowski_2003}%
  \BibitemOpen
  \bibfield  {author} {\bibinfo {author} {\bibfnamefont {W.}~\bibnamefont
  {Broniowski}}, \bibinfo {author} {\bibfnamefont {W.}~\bibnamefont
  {Florkowski}}, \ and\ \bibinfo {author} {\bibfnamefont {B.}~\bibnamefont
  {Hiller}},\ }\href {\doibase 10.1103/physrevc.68.034911} {\bibfield
  {journal} {\bibinfo  {journal} {Physical Review C}\ }\textbf {\bibinfo
  {volume} {68}} (\bibinfo {year} {2003}),\
  10.1103/physrevc.68.034911}\BibitemShut {NoStop}%
\bibitem [{\citenamefont {Ortega}\ \emph {et~al.}(2018)\citenamefont {Ortega},
  \citenamefont {Entem}, \citenamefont {Fernandez},\ and\ \citenamefont
  {Ruiz~Arriola}}]{Ortega:2017hpw}%
  \BibitemOpen
  \bibfield  {author} {\bibinfo {author} {\bibfnamefont {P.~G.}\ \bibnamefont
  {Ortega}}, \bibinfo {author} {\bibfnamefont {D.~R.}\ \bibnamefont {Entem}},
  \bibinfo {author} {\bibfnamefont {F.}~\bibnamefont {Fernandez}}, \ and\
  \bibinfo {author} {\bibfnamefont {E.}~\bibnamefont {Ruiz~Arriola}},\ }\href
  {\doibase 10.1016/j.physletb.2018.04.064} {\bibfield  {journal} {\bibinfo
  {journal} {Phys. Lett. B}\ }\textbf {\bibinfo {volume} {781}},\ \bibinfo
  {pages} {678} (\bibinfo {year} {2018})},\ \Eprint
  {http://arxiv.org/abs/1707.01915} {arXiv:1707.01915 [hep-ph]} \BibitemShut
  {NoStop}%
\bibitem [{\citenamefont {Lacroix}\ \emph {et~al.}(2015)\citenamefont
  {Lacroix}, \citenamefont {Semay},\ and\ \citenamefont
  {Buisseret}}]{Lacroix2015}%
  \BibitemOpen
  \bibfield  {author} {\bibinfo {author} {\bibfnamefont {G.}~\bibnamefont
  {Lacroix}}, \bibinfo {author} {\bibfnamefont {C.}~\bibnamefont {Semay}}, \
  and\ \bibinfo {author} {\bibfnamefont {F.}~\bibnamefont {Buisseret}},\ }\href
  {\doibase 10.1103/PhysRevC.91.065204} {\bibfield  {journal} {\bibinfo
  {journal} {Physical Review C - Nuclear Physics}\ } (\bibinfo {year} {2015}),\
  10.1103/PhysRevC.91.065204},\ \Eprint {http://arxiv.org/abs/1505.00540}
  {arXiv:1505.00540} \BibitemShut {NoStop}%
\bibitem [{\citenamefont {Huovinen}\ \emph {et~al.}(2017)\citenamefont
  {Huovinen}, \citenamefont {Lo}, \citenamefont {Marczenko}, \citenamefont
  {Morita}, \citenamefont {Redlich},\ and\ \citenamefont {Sasaki}}]{rho}%
  \BibitemOpen
  \bibfield  {author} {\bibinfo {author} {\bibfnamefont {P.}~\bibnamefont
  {Huovinen}}, \bibinfo {author} {\bibfnamefont {P.~M.}\ \bibnamefont {Lo}},
  \bibinfo {author} {\bibfnamefont {M.}~\bibnamefont {Marczenko}}, \bibinfo
  {author} {\bibfnamefont {K.}~\bibnamefont {Morita}}, \bibinfo {author}
  {\bibfnamefont {K.}~\bibnamefont {Redlich}}, \ and\ \bibinfo {author}
  {\bibfnamefont {C.}~\bibnamefont {Sasaki}},\ }\href {\doibase
  10.1016/j.physletb.2017.03.060} {\bibfield  {journal} {\bibinfo  {journal}
  {Physics Letters B}\ }\textbf {\bibinfo {volume} {769}},\ \bibinfo {pages}
  {509} (\bibinfo {year} {2017})},\ \Eprint {http://arxiv.org/abs/1608.06817}
  {arXiv:1608.06817 [hep-ph]} \BibitemShut {NoStop}%
\bibitem [{\citenamefont {Lo}(2018)}]{omega}%
  \BibitemOpen
  \bibfield  {author} {\bibinfo {author} {\bibfnamefont {P.~M.}\ \bibnamefont
  {Lo}},\ }\href {\doibase 10.1103/PhysRevC.97.035210} {\bibfield  {journal}
  {\bibinfo  {journal} {Phys. Rev. C}\ }\textbf {\bibinfo {volume} {97}},\
  \bibinfo {pages} {035210} (\bibinfo {year} {2018})},\ \Eprint
  {http://arxiv.org/abs/1705.01514} {arXiv:1705.01514 [hep-ph]} \BibitemShut
  {NoStop}%
\bibitem [{\citenamefont {Andronic}\ \emph
  {et~al.}(2018{\natexlab{a}})\citenamefont {Andronic}, \citenamefont
  {Braun-Munzinger}, \citenamefont {Friman}, \citenamefont {Lo}, \citenamefont
  {Redlich},\ and\ \citenamefont {Stachel}}]{ppuzz}%
  \BibitemOpen
  \bibfield  {author} {\bibinfo {author} {\bibfnamefont {A.}~\bibnamefont
  {Andronic}}, \bibinfo {author} {\bibfnamefont {P.}~\bibnamefont
  {Braun-Munzinger}}, \bibinfo {author} {\bibfnamefont {B.}~\bibnamefont
  {Friman}}, \bibinfo {author} {\bibfnamefont {P.~M.}\ \bibnamefont {Lo}},
  \bibinfo {author} {\bibfnamefont {K.}~\bibnamefont {Redlich}}, \ and\
  \bibinfo {author} {\bibfnamefont {J.}~\bibnamefont {Stachel}},\ }\href
  {\doibase 10.1016/j.physletb.2019.03.052} {\bibfield  {journal} {\bibinfo
  {journal} {Physics Letters B}\ }\textbf {\bibinfo {volume} {792}},\ \bibinfo
  {pages} {304} (\bibinfo {year} {2018}{\natexlab{a}})},\ \Eprint
  {http://arxiv.org/abs/1808.03102} {arXiv:1808.03102} \BibitemShut {NoStop}%
\bibitem [{\citenamefont {Horowitz}\ and\ \citenamefont
  {Schwenk}(2006{\natexlab{a}})}]{Horowitz:2005zv}%
  \BibitemOpen
  \bibfield  {author} {\bibinfo {author} {\bibfnamefont {C.}~\bibnamefont
  {Horowitz}}\ and\ \bibinfo {author} {\bibfnamefont {A.}~\bibnamefont
  {Schwenk}},\ }\href {\doibase 10.1016/j.physletb.2006.05.055} {\bibfield
  {journal} {\bibinfo  {journal} {Phys. Lett. B}\ }\textbf {\bibinfo {volume}
  {638}},\ \bibinfo {pages} {153} (\bibinfo {year} {2006}{\natexlab{a}})},\
  \Eprint {http://arxiv.org/abs/nucl-th/0507064} {arXiv:nucl-th/0507064}
  \BibitemShut {NoStop}%
\bibitem [{\citenamefont {Horowitz}\ and\ \citenamefont
  {Schwenk}(2006{\natexlab{b}})}]{Horowitz:2005nd}%
  \BibitemOpen
  \bibfield  {author} {\bibinfo {author} {\bibfnamefont {C.}~\bibnamefont
  {Horowitz}}\ and\ \bibinfo {author} {\bibfnamefont {A.}~\bibnamefont
  {Schwenk}},\ }\href {\doibase 10.1016/j.nuclphysa.2006.05.009} {\bibfield
  {journal} {\bibinfo  {journal} {Nucl. Phys. A}\ }\textbf {\bibinfo {volume}
  {776}},\ \bibinfo {pages} {55} (\bibinfo {year} {2006}{\natexlab{b}})},\
  \Eprint {http://arxiv.org/abs/nucl-th/0507033} {arXiv:nucl-th/0507033}
  \BibitemShut {NoStop}%
\bibitem [{\citenamefont {Mallik}\ \emph {et~al.}(2008)\citenamefont {Mallik},
  \citenamefont {De}, \citenamefont {Samaddar},\ and\ \citenamefont
  {Sarkar}}]{Mallik:2008zt}%
  \BibitemOpen
  \bibfield  {author} {\bibinfo {author} {\bibfnamefont {S.}~\bibnamefont
  {Mallik}}, \bibinfo {author} {\bibfnamefont {J.}~\bibnamefont {De}}, \bibinfo
  {author} {\bibfnamefont {S.}~\bibnamefont {Samaddar}}, \ and\ \bibinfo
  {author} {\bibfnamefont {S.}~\bibnamefont {Sarkar}},\ }\href {\doibase
  10.1103/PhysRevC.77.032201} {\bibfield  {journal} {\bibinfo  {journal} {Phys.
  Rev. C}\ }\textbf {\bibinfo {volume} {77}},\ \bibinfo {pages} {032201}
  (\bibinfo {year} {2008})},\ \Eprint {http://arxiv.org/abs/0801.0498}
  {arXiv:0801.0498 [nucl-th]} \BibitemShut {NoStop}%
\bibitem [{\citenamefont {Ropke}\ \emph {et~al.}(2013)\citenamefont {Ropke},
  \citenamefont {Bastian}, \citenamefont {Blaschke}, \citenamefont {Klahn},
  \citenamefont {Typel},\ and\ \citenamefont {Wolter}}]{Ropke:2012qv}%
  \BibitemOpen
  \bibfield  {author} {\bibinfo {author} {\bibfnamefont {G.}~\bibnamefont
  {Ropke}}, \bibinfo {author} {\bibfnamefont {N.-U.}\ \bibnamefont {Bastian}},
  \bibinfo {author} {\bibfnamefont {D.}~\bibnamefont {Blaschke}}, \bibinfo
  {author} {\bibfnamefont {T.}~\bibnamefont {Klahn}}, \bibinfo {author}
  {\bibfnamefont {S.}~\bibnamefont {Typel}}, \ and\ \bibinfo {author}
  {\bibfnamefont {H.}~\bibnamefont {Wolter}},\ }\href {\doibase
  10.1016/j.nuclphysa.2012.10.005} {\bibfield  {journal} {\bibinfo  {journal}
  {Nucl. Phys. A}\ }\textbf {\bibinfo {volume} {897}},\ \bibinfo {pages} {70}
  (\bibinfo {year} {2013})},\ \Eprint {http://arxiv.org/abs/1209.0212}
  {arXiv:1209.0212 [nucl-th]} \BibitemShut {NoStop}%
\bibitem [{\citenamefont {Oertel}\ \emph {et~al.}(2017)\citenamefont {Oertel},
  \citenamefont {Hempel}, \citenamefont {Klahn},\ and\ \citenamefont
  {Typel}}]{Oertel:2016bki}%
  \BibitemOpen
  \bibfield  {author} {\bibinfo {author} {\bibfnamefont {M.}~\bibnamefont
  {Oertel}}, \bibinfo {author} {\bibfnamefont {M.}~\bibnamefont {Hempel}},
  \bibinfo {author} {\bibfnamefont {T.}~\bibnamefont {Klahn}}, \ and\ \bibinfo
  {author} {\bibfnamefont {S.}~\bibnamefont {Typel}},\ }\href {\doibase
  10.1103/RevModPhys.89.015007} {\bibfield  {journal} {\bibinfo  {journal}
  {Rev. Mod. Phys.}\ }\textbf {\bibinfo {volume} {89}},\ \bibinfo {pages}
  {015007} (\bibinfo {year} {2017})},\ \Eprint
  {http://arxiv.org/abs/1610.03361} {arXiv:1610.03361 [astro-ph.HE]}
  \BibitemShut {NoStop}%
\bibitem [{\citenamefont {Fore}\ and\ \citenamefont {Reddy}(2020)}]{sanjay}%
  \BibitemOpen
  \bibfield  {author} {\bibinfo {author} {\bibfnamefont {B.}~\bibnamefont
  {Fore}}\ and\ \bibinfo {author} {\bibfnamefont {S.}~\bibnamefont {Reddy}},\
  }\href {\doibase 10.1103/PhysRevC.101.035809} {\bibfield  {journal} {\bibinfo
   {journal} {Phys. Rev. C}\ }\textbf {\bibinfo {volume} {101}},\ \bibinfo
  {pages} {035809} (\bibinfo {year} {2020})},\ \Eprint
  {http://arxiv.org/abs/1911.02632} {arXiv:1911.02632 [astro-ph.HE]}
  \BibitemShut {NoStop}%
\bibitem [{\citenamefont {Borsanyi}\ \emph {et~al.}(2012)\citenamefont
  {Borsanyi}, \citenamefont {Fodor}, \citenamefont {Katz}, \citenamefont
  {Krieg}, \citenamefont {Ratti},\ and\ \citenamefont
  {Szabo}}]{Borsanyi:2011sw}%
  \BibitemOpen
  \bibfield  {author} {\bibinfo {author} {\bibfnamefont {S.}~\bibnamefont
  {Borsanyi}}, \bibinfo {author} {\bibfnamefont {Z.}~\bibnamefont {Fodor}},
  \bibinfo {author} {\bibfnamefont {S.~D.}\ \bibnamefont {Katz}}, \bibinfo
  {author} {\bibfnamefont {S.}~\bibnamefont {Krieg}}, \bibinfo {author}
  {\bibfnamefont {C.}~\bibnamefont {Ratti}}, \ and\ \bibinfo {author}
  {\bibfnamefont {K.}~\bibnamefont {Szabo}},\ }\href {\doibase
  10.1007/JHEP01(2012)138} {\bibfield  {journal} {\bibinfo  {journal} {JHEP}\
  }\textbf {\bibinfo {volume} {01}},\ \bibinfo {pages} {138} (\bibinfo {year}
  {2012})},\ \Eprint {http://arxiv.org/abs/1112.4416} {arXiv:1112.4416
  [hep-lat]} \BibitemShut {NoStop}%
\bibitem [{\citenamefont {Bazavov}\ \emph
  {et~al.}(2012{\natexlab{a}})\citenamefont {Bazavov} \emph
  {et~al.}}]{Bazavov:2012jq}%
  \BibitemOpen
  \bibfield  {author} {\bibinfo {author} {\bibfnamefont {A.}~\bibnamefont
  {Bazavov}} \emph {et~al.} (\bibinfo {collaboration} {HotQCD}),\ }\href
  {\doibase 10.1103/PhysRevD.86.034509} {\bibfield  {journal} {\bibinfo
  {journal} {Phys. Rev. D}\ }\textbf {\bibinfo {volume} {86}},\ \bibinfo
  {pages} {034509} (\bibinfo {year} {2012}{\natexlab{a}})},\ \Eprint
  {http://arxiv.org/abs/1203.0784} {arXiv:1203.0784 [hep-lat]} \BibitemShut
  {NoStop}%
\bibitem [{\citenamefont {Huovinen}\ and\ \citenamefont
  {Petreczky}(2018)}]{Huovinen:2017ogf}%
  \BibitemOpen
  \bibfield  {author} {\bibinfo {author} {\bibfnamefont {P.}~\bibnamefont
  {Huovinen}}\ and\ \bibinfo {author} {\bibfnamefont {P.}~\bibnamefont
  {Petreczky}},\ }\href {\doibase 10.1016/j.physletb.2017.12.001} {\bibfield
  {journal} {\bibinfo  {journal} {Phys. Lett. B}\ }\textbf {\bibinfo {volume}
  {777}},\ \bibinfo {pages} {125} (\bibinfo {year} {2018})},\ \Eprint
  {http://arxiv.org/abs/1708.00879} {arXiv:1708.00879 [hep-ph]} \BibitemShut
  {NoStop}%
\bibitem [{\citenamefont {Dash}\ \emph {et~al.}(2018)\citenamefont {Dash},
  \citenamefont {Samanta},\ and\ \citenamefont {Mohanty}}]{kmat}%
  \BibitemOpen
  \bibfield  {author} {\bibinfo {author} {\bibfnamefont {A.}~\bibnamefont
  {Dash}}, \bibinfo {author} {\bibfnamefont {S.}~\bibnamefont {Samanta}}, \
  and\ \bibinfo {author} {\bibfnamefont {B.}~\bibnamefont {Mohanty}},\ }\href
  {https://arxiv.org/pdf/1802.04998} {\bibfield  {journal} {\bibinfo  {journal}
  {Phys. Rev. C}\ }\textbf {\bibinfo {volume} {97}},\ \bibinfo {pages} {055208}
  (\bibinfo {year} {2018})},\ \Eprint {http://arxiv.org/abs/1802.04998}
  {1802.04998} \BibitemShut {NoStop}%
\bibitem [{\citenamefont {Friman}\ \emph {et~al.}(2015)\citenamefont {Friman},
  \citenamefont {Lo}, \citenamefont {Marczenko}, \citenamefont {Redlich},\ and\
  \citenamefont {Sasaki}}]{kappa}%
  \BibitemOpen
  \bibfield  {author} {\bibinfo {author} {\bibfnamefont {B.}~\bibnamefont
  {Friman}}, \bibinfo {author} {\bibfnamefont {P.~M.}\ \bibnamefont {Lo}},
  \bibinfo {author} {\bibfnamefont {M.}~\bibnamefont {Marczenko}}, \bibinfo
  {author} {\bibfnamefont {K.}~\bibnamefont {Redlich}}, \ and\ \bibinfo
  {author} {\bibfnamefont {C.}~\bibnamefont {Sasaki}},\ }\href {\doibase
  10.1103/PhysRevD.92.074003} {\bibfield  {journal} {\bibinfo  {journal} {Phys.
  Rev. D}\ }\textbf {\bibinfo {volume} {92}},\ \bibinfo {pages} {074003}
  (\bibinfo {year} {2015})},\ \Eprint {http://arxiv.org/abs/1507.04183}
  {arXiv:1507.04183 [hep-ph]} \BibitemShut {NoStop}%
\bibitem [{\citenamefont {Lo}\ \emph {et~al.}(2018)\citenamefont {Lo},
  \citenamefont {Friman}, \citenamefont {Redlich},\ and\ \citenamefont
  {Sasaki}}]{chibq}%
  \BibitemOpen
  \bibfield  {author} {\bibinfo {author} {\bibfnamefont {P.~M.}\ \bibnamefont
  {Lo}}, \bibinfo {author} {\bibfnamefont {B.}~\bibnamefont {Friman}}, \bibinfo
  {author} {\bibfnamefont {K.}~\bibnamefont {Redlich}}, \ and\ \bibinfo
  {author} {\bibfnamefont {C.}~\bibnamefont {Sasaki}},\ }\href {\doibase
  10.1016/j.physletb.2018.01.016} {\bibfield  {journal} {\bibinfo  {journal}
  {Phys. Lett. B}\ }\textbf {\bibinfo {volume} {778}},\ \bibinfo {pages} {454}
  (\bibinfo {year} {2018})},\ \Eprint {http://arxiv.org/abs/1710.02711}
  {arXiv:1710.02711 [hep-ph]} \BibitemShut {NoStop}%
\bibitem [{\citenamefont {Fern\'andez-Ram\'{\i}rez}\ \emph
  {et~al.}(2018)\citenamefont {Fern\'andez-Ram\'{\i}rez}, \citenamefont {Lo},\
  and\ \citenamefont {Petreczky}}]{chibs}%
  \BibitemOpen
  \bibfield  {author} {\bibinfo {author} {\bibfnamefont {C.}~\bibnamefont
  {Fern\'andez-Ram\'{\i}rez}}, \bibinfo {author} {\bibfnamefont {P.~M.}\
  \bibnamefont {Lo}}, \ and\ \bibinfo {author} {\bibfnamefont {P.}~\bibnamefont
  {Petreczky}},\ }\href {\doibase 10.1103/PhysRevC.98.044910} {\bibfield
  {journal} {\bibinfo  {journal} {Phys. Rev. C}\ }\textbf {\bibinfo {volume}
  {98}},\ \bibinfo {pages} {044910} (\bibinfo {year} {2018})}\BibitemShut
  {NoStop}%
\bibitem [{\citenamefont {Taylor}(2012)}]{taylor}%
  \BibitemOpen
  \bibfield  {author} {\bibinfo {author} {\bibfnamefont {J.}~\bibnamefont
  {Taylor}},\ }\href {https://books.google.pl/books?id=OIaXvuwZMLQC} {\emph
  {\bibinfo {title} {Scattering Theory: The Quantum Theory of Nonrelativistic
  Collisions}}},\ Dover Books on Engineering\ (\bibinfo  {publisher} {Dover
  Publications},\ \bibinfo {year} {2012})\BibitemShut {NoStop}%
\bibitem [{\citenamefont {Dashen}\ and\ \citenamefont
  {Rajaraman}(1974)}]{elementarity}%
  \BibitemOpen
  \bibfield  {author} {\bibinfo {author} {\bibfnamefont {R.}~\bibnamefont
  {Dashen}}\ and\ \bibinfo {author} {\bibfnamefont {R.}~\bibnamefont
  {Rajaraman}},\ }\href {\doibase 10.1103/PhysRevD.10.708} {\bibfield
  {journal} {\bibinfo  {journal} {Phys. Rev. D}\ }\textbf {\bibinfo {volume}
  {10}},\ \bibinfo {pages} {708} (\bibinfo {year} {1974})}\BibitemShut
  {NoStop}%
\bibitem [{\citenamefont {Weinhold}\ \emph {et~al.}(1998)\citenamefont
  {Weinhold}, \citenamefont {Friman},\ and\ \citenamefont
  {Norenberg}}]{weinhold}%
  \BibitemOpen
  \bibfield  {author} {\bibinfo {author} {\bibfnamefont {W.}~\bibnamefont
  {Weinhold}}, \bibinfo {author} {\bibfnamefont {B.}~\bibnamefont {Friman}}, \
  and\ \bibinfo {author} {\bibfnamefont {W.}~\bibnamefont {Norenberg}},\ }\href
  {\doibase 10.1016/S0370-2693(98)00639-X} {\bibfield  {journal} {\bibinfo
  {journal} {Phys. Lett. B}\ }\textbf {\bibinfo {volume} {433}},\ \bibinfo
  {pages} {236} (\bibinfo {year} {1998})},\ \Eprint
  {http://arxiv.org/abs/nucl-th/9710014} {arXiv:nucl-th/9710014 [nucl-th]}
  \BibitemShut {NoStop}%
\bibitem [{\citenamefont {Lo}\ and\ \citenamefont {Giacosa}(2019)}]{lee}%
  \BibitemOpen
  \bibfield  {author} {\bibinfo {author} {\bibfnamefont {P.~M.}\ \bibnamefont
  {Lo}}\ and\ \bibinfo {author} {\bibfnamefont {F.}~\bibnamefont {Giacosa}},\
  }\href {\doibase 10.1140/epjc/s10052-019-6844-y} {\bibfield  {journal}
  {\bibinfo  {journal} {The European Physical Journal C}\ }\textbf {\bibinfo
  {volume} {79}} (\bibinfo {year} {2019}),\
  10.1140/epjc/s10052-019-6844-y}\BibitemShut {NoStop}%
\bibitem [{\citenamefont {{Giacosa, Francesco}}(2020)}]{Giacosa:2020tha}%
  \BibitemOpen
  \bibfield  {author} {\bibinfo {author} {\bibnamefont {{Giacosa,
  Francesco}}},\ }\href@noop {} {\  (\bibinfo {year} {2020})},\ \Eprint
  {http://arxiv.org/abs/2001.07781} {arXiv:2001.07781 [hep-ph]} \BibitemShut
  {NoStop}%
\bibitem [{\citenamefont {Weidenm{\"u}ller}(1967)}]{weidenmueller}%
  \BibitemOpen
  \bibfield  {author} {\bibinfo {author} {\bibfnamefont {H.}~\bibnamefont
  {Weidenm{\"u}ller}},\ }\href {\doibase
  https://doi.org/10.1016/0370-2693(67)90264-X} {\bibfield  {journal} {\bibinfo
   {journal} {Physics Letters B}\ }\textbf {\bibinfo {volume} {24}},\ \bibinfo
  {pages} {441 } (\bibinfo {year} {1967})}\BibitemShut {NoStop}%
\bibitem [{\citenamefont {Gerber}\ and\ \citenamefont
  {Leutwyler}(1989)}]{chipt}%
  \BibitemOpen
  \bibfield  {author} {\bibinfo {author} {\bibfnamefont {P.}~\bibnamefont
  {Gerber}}\ and\ \bibinfo {author} {\bibfnamefont {H.}~\bibnamefont
  {Leutwyler}},\ }\href {\doibase https://doi.org/10.1016/0550-3213(89)90349-0}
  {\bibfield  {journal} {\bibinfo  {journal} {Nuclear Physics B}\ }\textbf
  {\bibinfo {volume} {321}},\ \bibinfo {pages} {387 } (\bibinfo {year}
  {1989})}\BibitemShut {NoStop}%
\bibitem [{\citenamefont {Oller}\ \emph {et~al.}(1999)\citenamefont {Oller},
  \citenamefont {Oset},\ and\ \citenamefont {Pel\'aez}}]{Oller:2007aa}%
  \BibitemOpen
  \bibfield  {author} {\bibinfo {author} {\bibfnamefont {J.~A.}\ \bibnamefont
  {Oller}}, \bibinfo {author} {\bibfnamefont {E.}~\bibnamefont {Oset}}, \ and\
  \bibinfo {author} {\bibfnamefont {J.~R.}\ \bibnamefont {Pel\'aez}},\ }\href
  {\doibase 10.1103/PhysRevD.59.074001} {\bibfield  {journal} {\bibinfo
  {journal} {Phys. Rev. D}\ }\textbf {\bibinfo {volume} {59}},\ \bibinfo
  {pages} {074001} (\bibinfo {year} {1999})}\BibitemShut {NoStop}%
\bibitem [{\citenamefont {Hagedorn}(1965)}]{Hagedorn:1965st}%
  \BibitemOpen
  \bibfield  {author} {\bibinfo {author} {\bibfnamefont {R.}~\bibnamefont
  {Hagedorn}},\ }\href@noop {} {\bibfield  {journal} {\bibinfo  {journal}
  {Nuovo Cim. Suppl.}\ }\textbf {\bibinfo {volume} {3}},\ \bibinfo {pages}
  {147} (\bibinfo {year} {1965})}\BibitemShut {NoStop}%
\bibitem [{\citenamefont {Andronic}\ \emph
  {et~al.}(2018{\natexlab{b}})\citenamefont {Andronic}, \citenamefont
  {Braun-Munzinger}, \citenamefont {Redlich},\ and\ \citenamefont
  {Stachel}}]{hrgnature}%
  \BibitemOpen
  \bibfield  {author} {\bibinfo {author} {\bibfnamefont {A.}~\bibnamefont
  {Andronic}}, \bibinfo {author} {\bibfnamefont {P.}~\bibnamefont
  {Braun-Munzinger}}, \bibinfo {author} {\bibfnamefont {K.}~\bibnamefont
  {Redlich}}, \ and\ \bibinfo {author} {\bibfnamefont {J.}~\bibnamefont
  {Stachel}},\ }\href {\doibase 10.1038/s41586-018-0491-6} {\bibfield
  {journal} {\bibinfo  {journal} {Nature}\ }\textbf {\bibinfo {volume} {561}},\
  \bibinfo {pages} {321} (\bibinfo {year} {2018}{\natexlab{b}})},\ \Eprint
  {http://arxiv.org/abs/1710.09425} {arXiv:1710.09425 [nucl-th]} \BibitemShut
  {NoStop}%
\bibitem [{Note1()}]{Note1}%
  \BibitemOpen
  \bibinfo {note} {Note that ${\protect \rm Im} \protect \tmspace +\thinmuskip
  {.1667em} \protect \qopname \relax o{ln}(m_{\protect \rm res} - E \pm i
  \protect \tmspace +\thinmuskip {.1667em} 0^+ ) = \pm \pi \protect \tmspace
  +\thinmuskip {.1667em} \theta (E-m_{\protect \rm res})$. It is recommended to
  use {\protect \it atan2} numerical implementation (available in most
  programming language, e.g. c++, fortran, and python) to extract the phase,
  where this relation is automatic.}\BibitemShut {Stop}%
\bibitem [{\citenamefont {Colangelo}\ \emph {et~al.}(2001)\citenamefont
  {Colangelo}, \citenamefont {Gasser},\ and\ \citenamefont
  {Leutwyler}}]{COLANGELO2001125}%
  \BibitemOpen
  \bibfield  {author} {\bibinfo {author} {\bibfnamefont {G.}~\bibnamefont
  {Colangelo}}, \bibinfo {author} {\bibfnamefont {J.}~\bibnamefont {Gasser}}, \
  and\ \bibinfo {author} {\bibfnamefont {H.}~\bibnamefont {Leutwyler}},\ }\href
  {\doibase https://doi.org/10.1016/S0550-3213(01)00147-X} {\bibfield
  {journal} {\bibinfo  {journal} {Nuclear Physics B}\ }\textbf {\bibinfo
  {volume} {603}},\ \bibinfo {pages} {125 } (\bibinfo {year}
  {2001})}\BibitemShut {NoStop}%
\bibitem [{\citenamefont {Garc\'{\i}a-Mart\'{\i}n}\ \emph
  {et~al.}(2011)\citenamefont {Garc\'{\i}a-Mart\'{\i}n}, \citenamefont
  {Kami\ifmmode~\acute{n}\else \'{n}\fi{}ski}, \citenamefont {Pel\'aez},
  \citenamefont {Ruiz~de Elvira},\ and\ \citenamefont
  {Yndur\'ain}}]{PhysRevD.83.074004}%
  \BibitemOpen
  \bibfield  {author} {\bibinfo {author} {\bibfnamefont {R.}~\bibnamefont
  {Garc\'{\i}a-Mart\'{\i}n}}, \bibinfo {author} {\bibfnamefont
  {R.}~\bibnamefont {Kami\ifmmode~\acute{n}\else \'{n}\fi{}ski}}, \bibinfo
  {author} {\bibfnamefont {J.~R.}\ \bibnamefont {Pel\'aez}}, \bibinfo {author}
  {\bibfnamefont {J.}~\bibnamefont {Ruiz~de Elvira}}, \ and\ \bibinfo {author}
  {\bibfnamefont {F.~J.}\ \bibnamefont {Yndur\'ain}},\ }\href {\doibase
  10.1103/PhysRevD.83.074004} {\bibfield  {journal} {\bibinfo  {journal} {Phys.
  Rev. D}\ }\textbf {\bibinfo {volume} {83}},\ \bibinfo {pages} {074004}
  (\bibinfo {year} {2011})}\BibitemShut {NoStop}%
\bibitem [{\citenamefont {Markushin}(2000)}]{Markushin:2000fa}%
  \BibitemOpen
  \bibfield  {author} {\bibinfo {author} {\bibfnamefont {V.}~\bibnamefont
  {Markushin}},\ }\href {\doibase 10.1007/s100500070092} {\bibfield  {journal}
  {\bibinfo  {journal} {Eur. Phys. J. A}\ }\textbf {\bibinfo {volume} {8}},\
  \bibinfo {pages} {389} (\bibinfo {year} {2000})},\ \Eprint
  {http://arxiv.org/abs/hep-ph/0005164} {arXiv:hep-ph/0005164} \BibitemShut
  {NoStop}%
\bibitem [{\citenamefont {Locher}\ \emph {et~al.}(1998)\citenamefont {Locher},
  \citenamefont {Markushin},\ and\ \citenamefont {Zheng}}]{Locher:1997gr}%
  \BibitemOpen
  \bibfield  {author} {\bibinfo {author} {\bibfnamefont {M.}~\bibnamefont
  {Locher}}, \bibinfo {author} {\bibfnamefont {V.}~\bibnamefont {Markushin}}, \
  and\ \bibinfo {author} {\bibfnamefont {H.}~\bibnamefont {Zheng}},\ }\href
  {\doibase 10.1007/s100520050210} {\bibfield  {journal} {\bibinfo  {journal}
  {Eur. Phys. J. C}\ }\textbf {\bibinfo {volume} {4}},\ \bibinfo {pages} {317}
  (\bibinfo {year} {1998})},\ \Eprint {http://arxiv.org/abs/hep-ph/9705230}
  {arXiv:hep-ph/9705230} \BibitemShut {NoStop}%
\bibitem [{\citenamefont {Morgan}\ and\ \citenamefont
  {Pennington}(1993)}]{Morgan:1993td}%
  \BibitemOpen
  \bibfield  {author} {\bibinfo {author} {\bibfnamefont {D.}~\bibnamefont
  {Morgan}}\ and\ \bibinfo {author} {\bibfnamefont {M.}~\bibnamefont
  {Pennington}},\ }\href {\doibase 10.1103/PhysRevD.48.1185} {\bibfield
  {journal} {\bibinfo  {journal} {Phys. Rev. D}\ }\textbf {\bibinfo {volume}
  {48}},\ \bibinfo {pages} {1185} (\bibinfo {year} {1993})}\BibitemShut
  {NoStop}%
\bibitem [{\citenamefont {Donoghue}(1996)}]{Donoghue:1996kw}%
  \BibitemOpen
  \bibfield  {author} {\bibinfo {author} {\bibfnamefont {J.~F.}\ \bibnamefont
  {Donoghue}},\ }in\ \href@noop {} {\emph {\bibinfo {booktitle} {{Advanced
  School on Effective Theories}}}}\ (\bibinfo {year} {1996})\ \Eprint
  {http://arxiv.org/abs/hep-ph/9607351} {arXiv:hep-ph/9607351} \BibitemShut
  {NoStop}%
\bibitem [{\citenamefont {Zwicky}(2016)}]{Zwicky2016}%
  \BibitemOpen
  \bibfield  {author} {\bibinfo {author} {\bibfnamefont {R.}~\bibnamefont
  {Zwicky}},\ }\href {http://arxiv.org/abs/1610.06090} {\  (\bibinfo {year}
  {2016})},\ \Eprint {http://arxiv.org/abs/1610.06090} {arXiv:1610.06090}
  \BibitemShut {NoStop}%
\bibitem [{Note2()}]{Note2}%
  \BibitemOpen
  \bibinfo {note} {Here we need to follow the convention of Ref.~\cite
  {Markushin:2000fa} to use their fitting parameters. Checking for the diagonal
  element and note that ${\protect \rm Im} \protect \tmspace +\thinmuskip
  {.1667em} G^0_\alpha (s) \sim -q_\alpha $, we see that $ S_{\alpha \alpha }
  \sim 1 - 2 \protect \tmspace +\thinmuskip {.1667em} i \protect \tmspace
  +\thinmuskip {.1667em} T_{\alpha \alpha }$, identifying $T_{\alpha \alpha } =
  - f_\alpha $ to the scattering amplitude, instead of the standard $T_{\alpha
  \alpha } = -8 \protect \tmspace +\thinmuskip {.1667em} \pi \protect \tmspace
  +\thinmuskip {.1667em} \protect \sqrt {s} \protect \tmspace +\thinmuskip
  {.1667em} f_\alpha $.}\BibitemShut {Stop}%
\bibitem [{\citenamefont {Churchill}\ and\ \citenamefont
  {Brown}(2009)}]{Churchill2009}%
  \BibitemOpen
  \bibfield  {author} {\bibinfo {author} {\bibfnamefont {R.~V.}\ \bibnamefont
  {Churchill}}\ and\ \bibinfo {author} {\bibfnamefont {J.~W.}\ \bibnamefont
  {Brown}},\ }\href@noop {} {\emph {\bibinfo {title} {Complex Variables}}}\
  (\bibinfo {year} {2009})\ p.\ \bibinfo {pages} {499}\BibitemShut {NoStop}%
\bibitem [{\citenamefont {Wegert}(2012)}]{wegert2012visual}%
  \BibitemOpen
  \bibfield  {author} {\bibinfo {author} {\bibfnamefont {E.}~\bibnamefont
  {Wegert}},\ }\href {https://books.google.pl/books?id=zRppM7WO0vIC} {\emph
  {\bibinfo {title} {Visual Complex Functions: An Introduction with Phase
  Portraits}}},\ SpringerLink : B{\"u}cher\ (\bibinfo  {publisher} {Springer
  Basel},\ \bibinfo {year} {2012})\BibitemShut {NoStop}%
\bibitem [{\citenamefont {Kaminski}\ \emph {et~al.}(1999)\citenamefont
  {Kaminski}, \citenamefont {Lesniak},\ and\ \citenamefont
  {Loiseau}}]{Kaminski:1998ns}%
  \BibitemOpen
  \bibfield  {author} {\bibinfo {author} {\bibfnamefont {R.}~\bibnamefont
  {Kaminski}}, \bibinfo {author} {\bibfnamefont {L.}~\bibnamefont {Lesniak}}, \
  and\ \bibinfo {author} {\bibfnamefont {B.}~\bibnamefont {Loiseau}},\ }\href
  {\doibase 10.1007/s100529900023} {\bibfield  {journal} {\bibinfo  {journal}
  {Eur. Phys. J. C}\ }\textbf {\bibinfo {volume} {9}},\ \bibinfo {pages} {141}
  (\bibinfo {year} {1999})},\ \Eprint {http://arxiv.org/abs/hep-ph/9810386}
  {arXiv:hep-ph/9810386} \BibitemShut {NoStop}%
\bibitem [{\citenamefont {Kaminski}\ \emph {et~al.}(1994)\citenamefont
  {Kaminski}, \citenamefont {Lesniak},\ and\ \citenamefont
  {Maillet}}]{Kaminski:1993zb}%
  \BibitemOpen
  \bibfield  {author} {\bibinfo {author} {\bibfnamefont {R.}~\bibnamefont
  {Kaminski}}, \bibinfo {author} {\bibfnamefont {L.}~\bibnamefont {Lesniak}}, \
  and\ \bibinfo {author} {\bibfnamefont {J.}~\bibnamefont {Maillet}},\ }\href
  {\doibase 10.1103/PhysRevD.50.3145} {\bibfield  {journal} {\bibinfo
  {journal} {Phys. Rev. D}\ }\textbf {\bibinfo {volume} {50}},\ \bibinfo
  {pages} {3145} (\bibinfo {year} {1994})},\ \Eprint
  {http://arxiv.org/abs/hep-ph/9403264} {arXiv:hep-ph/9403264} \BibitemShut
  {NoStop}%
\bibitem [{\citenamefont {Frazer}\ and\ \citenamefont
  {Hendry}(1964)}]{Frazer:1964zz}%
  \BibitemOpen
  \bibfield  {author} {\bibinfo {author} {\bibfnamefont {W.~R.}\ \bibnamefont
  {Frazer}}\ and\ \bibinfo {author} {\bibfnamefont {A.~W.}\ \bibnamefont
  {Hendry}},\ }\href {\doibase 10.1103/PhysRev.134.B1307} {\bibfield  {journal}
  {\bibinfo  {journal} {Phys. Rev.}\ }\textbf {\bibinfo {volume} {134}},\
  \bibinfo {pages} {B1307} (\bibinfo {year} {1964})}\BibitemShut {NoStop}%
\bibitem [{\citenamefont {Badalyan}\ \emph {et~al.}(1982)\citenamefont
  {Badalyan}, \citenamefont {Kok}, \citenamefont {Polikarpov},\ and\
  \citenamefont {Simonov}}]{riemann_shts}%
  \BibitemOpen
  \bibfield  {author} {\bibinfo {author} {\bibfnamefont {A.}~\bibnamefont
  {Badalyan}}, \bibinfo {author} {\bibfnamefont {L.}~\bibnamefont {Kok}},
  \bibinfo {author} {\bibfnamefont {M.}~\bibnamefont {Polikarpov}}, \ and\
  \bibinfo {author} {\bibfnamefont {Y.}~\bibnamefont {Simonov}},\ }\href
  {\doibase https://doi.org/10.1016/0370-1573(82)90014-X} {\bibfield  {journal}
  {\bibinfo  {journal} {Physics Reports}\ }\textbf {\bibinfo {volume} {82}},\
  \bibinfo {pages} {31 } (\bibinfo {year} {1982})}\BibitemShut {NoStop}%
\bibitem [{\citenamefont {Broniowski}\ \emph {et~al.}(2015)\citenamefont
  {Broniowski}, \citenamefont {Giacosa},\ and\ \citenamefont {Begun}}]{sigma}%
  \BibitemOpen
  \bibfield  {author} {\bibinfo {author} {\bibfnamefont {W.}~\bibnamefont
  {Broniowski}}, \bibinfo {author} {\bibfnamefont {F.}~\bibnamefont {Giacosa}},
  \ and\ \bibinfo {author} {\bibfnamefont {V.}~\bibnamefont {Begun}},\ }\href
  {\doibase 10.1103/PhysRevC.92.034905} {\bibfield  {journal} {\bibinfo
  {journal} {Phys. Rev. C}\ }\textbf {\bibinfo {volume} {92}},\ \bibinfo
  {pages} {034905} (\bibinfo {year} {2015})},\ \Eprint
  {http://arxiv.org/abs/1506.01260} {arXiv:1506.01260 [nucl-th]} \BibitemShut
  {NoStop}%
\bibitem [{\citenamefont {Vovchenko}\ \emph {et~al.}(2018)\citenamefont
  {Vovchenko}, \citenamefont {Motornenko}, \citenamefont {Gorenstein},\ and\
  \citenamefont {Stoecker}}]{Vovchenko:2017drx}%
  \BibitemOpen
  \bibfield  {author} {\bibinfo {author} {\bibfnamefont {V.}~\bibnamefont
  {Vovchenko}}, \bibinfo {author} {\bibfnamefont {A.}~\bibnamefont
  {Motornenko}}, \bibinfo {author} {\bibfnamefont {M.~I.}\ \bibnamefont
  {Gorenstein}}, \ and\ \bibinfo {author} {\bibfnamefont {H.}~\bibnamefont
  {Stoecker}},\ }\href {\doibase 10.1103/PhysRevC.97.035202} {\bibfield
  {journal} {\bibinfo  {journal} {Phys. Rev. C}\ }\textbf {\bibinfo {volume}
  {97}},\ \bibinfo {pages} {035202} (\bibinfo {year} {2018})},\ \Eprint
  {http://arxiv.org/abs/1710.00693} {arXiv:1710.00693 [nucl-th]} \BibitemShut
  {NoStop}%
\bibitem [{\citenamefont {Lo}\ \emph {et~al.}(2017)\citenamefont {Lo},
  \citenamefont {Friman}, \citenamefont {Marczenko}, \citenamefont {Redlich},\
  and\ \citenamefont {Sasaki}}]{exclvol}%
  \BibitemOpen
  \bibfield  {author} {\bibinfo {author} {\bibfnamefont {P.~M.}\ \bibnamefont
  {Lo}}, \bibinfo {author} {\bibfnamefont {B.}~\bibnamefont {Friman}}, \bibinfo
  {author} {\bibfnamefont {M.}~\bibnamefont {Marczenko}}, \bibinfo {author}
  {\bibfnamefont {K.}~\bibnamefont {Redlich}}, \ and\ \bibinfo {author}
  {\bibfnamefont {C.}~\bibnamefont {Sasaki}},\ }\href {\doibase
  10.1103/PhysRevC.96.015207} {\bibfield  {journal} {\bibinfo  {journal} {Phys.
  Rev. C}\ }\textbf {\bibinfo {volume} {96}},\ \bibinfo {pages} {015207}
  (\bibinfo {year} {2017})},\ \Eprint {http://arxiv.org/abs/1703.00306}
  {arXiv:1703.00306 [nucl-th]} \BibitemShut {NoStop}%
\bibitem [{Note3()}]{Note3}%
  \BibitemOpen
  \bibinfo {note} {We expect this point, and the arrangement of poles and roots
  in general, to be model dependent, and can depend on the prescription (Eq.
  (\ref {eq:smat})) in extracting the complex landscape of S-matrix. The result
  on the physical line, on the other hand, is not expected to
  change.}\BibitemShut {Stop}%
\bibitem [{\citenamefont {Kami\'nski}\ and\ \citenamefont
  {Bochnacki}(2019)}]{Kaminski:2019bep}%
  \BibitemOpen
  \bibfield  {author} {\bibinfo {author} {\bibfnamefont {R.}~\bibnamefont
  {Kami\'nski}}\ and\ \bibinfo {author} {\bibfnamefont {P.}~\bibnamefont
  {Bochnacki}},\ }\href {\doibase 10.5506/APhysPolB.50.1911} {\bibfield
  {journal} {\bibinfo  {journal} {Acta Phys. Polon. B}\ }\textbf {\bibinfo
  {volume} {50}},\ \bibinfo {pages} {1911} (\bibinfo {year}
  {2019})}\BibitemShut {NoStop}%
\bibitem [{\citenamefont {Swanson}(2015)}]{cusp}%
  \BibitemOpen
  \bibfield  {author} {\bibinfo {author} {\bibfnamefont {E.~S.}\ \bibnamefont
  {Swanson}},\ }\href {\doibase 10.1103/PhysRevD.91.034009} {\bibfield
  {journal} {\bibinfo  {journal} {Phys. Rev. D}\ }\textbf {\bibinfo {volume}
  {91}},\ \bibinfo {pages} {034009} (\bibinfo {year} {2015})}\BibitemShut
  {NoStop}%
\bibitem [{\citenamefont {Mikhasenko}\ \emph {et~al.}(2015)\citenamefont
  {Mikhasenko}, \citenamefont {Ketzer},\ and\ \citenamefont
  {Sarantsev}}]{triangle}%
  \BibitemOpen
  \bibfield  {author} {\bibinfo {author} {\bibfnamefont {M.}~\bibnamefont
  {Mikhasenko}}, \bibinfo {author} {\bibfnamefont {B.}~\bibnamefont {Ketzer}},
  \ and\ \bibinfo {author} {\bibfnamefont {A.}~\bibnamefont {Sarantsev}},\
  }\href {\doibase 10.1103/PhysRevD.91.094015} {\bibfield  {journal} {\bibinfo
  {journal} {Phys. Rev. D}\ }\textbf {\bibinfo {volume} {91}},\ \bibinfo
  {pages} {094015} (\bibinfo {year} {2015})}\BibitemShut {NoStop}%
\bibitem [{\citenamefont {Briceno}\ \emph {et~al.}(2018)\citenamefont
  {Briceno}, \citenamefont {Dudek},\ and\ \citenamefont {Young}}]{LQCD}%
  \BibitemOpen
  \bibfield  {author} {\bibinfo {author} {\bibfnamefont {R.~A.}\ \bibnamefont
  {Briceno}}, \bibinfo {author} {\bibfnamefont {J.~J.}\ \bibnamefont {Dudek}},
  \ and\ \bibinfo {author} {\bibfnamefont {R.~D.}\ \bibnamefont {Young}},\
  }\href {\doibase 10.1103/RevModPhys.90.025001} {\bibfield  {journal}
  {\bibinfo  {journal} {Rev. Mod. Phys.}\ }\textbf {\bibinfo {volume} {90}},\
  \bibinfo {pages} {025001} (\bibinfo {year} {2018})},\ \Eprint
  {http://arxiv.org/abs/1706.06223} {arXiv:1706.06223 [hep-lat]} \BibitemShut
  {NoStop}%
\bibitem [{\citenamefont {Bazavov}\ \emph {et~al.}(2014)\citenamefont {Bazavov}
  \emph {et~al.}}]{Bazavov:2014xya}%
  \BibitemOpen
  \bibfield  {author} {\bibinfo {author} {\bibfnamefont {A.}~\bibnamefont
  {Bazavov}} \emph {et~al.},\ }\href {\doibase 10.1103/PhysRevLett.113.072001}
  {\bibfield  {journal} {\bibinfo  {journal} {Phys. Rev. Lett.}\ }\textbf
  {\bibinfo {volume} {113}},\ \bibinfo {pages} {072001} (\bibinfo {year}
  {2014})},\ \Eprint {http://arxiv.org/abs/1404.6511} {arXiv:1404.6511
  [hep-lat]} \BibitemShut {NoStop}%
\bibitem [{\citenamefont {Lo}\ \emph {et~al.}(2015)\citenamefont {Lo},
  \citenamefont {Marczenko}, \citenamefont {Redlich},\ and\ \citenamefont
  {Sasaki}}]{missS}%
  \BibitemOpen
  \bibfield  {author} {\bibinfo {author} {\bibfnamefont {P.~M.}\ \bibnamefont
  {Lo}}, \bibinfo {author} {\bibfnamefont {M.}~\bibnamefont {Marczenko}},
  \bibinfo {author} {\bibfnamefont {K.}~\bibnamefont {Redlich}}, \ and\
  \bibinfo {author} {\bibfnamefont {C.}~\bibnamefont {Sasaki}},\ }\href
  {\doibase 10.1103/PhysRevC.92.055206} {\bibfield  {journal} {\bibinfo
  {journal} {Physical Review C}\ }\textbf {\bibinfo {volume} {92}},\ \bibinfo
  {pages} {055206} (\bibinfo {year} {2015})},\ \Eprint
  {http://arxiv.org/abs/1507.06398} {arXiv:1507.06398 [nucl-th]} \BibitemShut
  {NoStop}%
\bibitem [{\citenamefont {Swanson}(2005)}]{Swanson_2005}%
  \BibitemOpen
  \bibfield  {author} {\bibinfo {author} {\bibfnamefont {E.~S.}\ \bibnamefont
  {Swanson}},\ }\href {\doibase 10.1088/0954-3899/31/7/025} {\bibfield
  {journal} {\bibinfo  {journal} {Journal of Physics G: Nuclear and Particle
  Physics}\ }\textbf {\bibinfo {volume} {31}},\ \bibinfo {pages} {845–854}
  (\bibinfo {year} {2005})}\BibitemShut {NoStop}%
\bibitem [{\citenamefont {Coito}(2014)}]{Coito:2014dya}%
  \BibitemOpen
  \bibfield  {author} {\bibinfo {author} {\bibfnamefont {S.~P.~S.}\
  \bibnamefont {Coito}},\ }\href@noop {} {\  (\bibinfo {year} {2014})},\
  \Eprint {http://arxiv.org/abs/1401.7856} {arXiv:1401.7856 [hep-ph]}
  \BibitemShut {NoStop}%
\bibitem [{\citenamefont {Rupp}\ \emph {et~al.}(2015)\citenamefont {Rupp},
  \citenamefont {van Beveren},\ and\ \citenamefont {Coito}}]{Rupp:2015taa}%
  \BibitemOpen
  \bibfield  {author} {\bibinfo {author} {\bibfnamefont {G.}~\bibnamefont
  {Rupp}}, \bibinfo {author} {\bibfnamefont {E.}~\bibnamefont {van Beveren}}, \
  and\ \bibinfo {author} {\bibfnamefont {S.}~\bibnamefont {Coito}},\ }\href
  {\doibase 10.5506/APhysPolBSupp.8.139} {\bibfield  {journal} {\bibinfo
  {journal} {Acta Phys. Polon. Supp.}\ }\textbf {\bibinfo {volume} {8}},\
  \bibinfo {pages} {139} (\bibinfo {year} {2015})},\ \Eprint
  {http://arxiv.org/abs/1502.05250} {arXiv:1502.05250 [hep-ph]} \BibitemShut
  {NoStop}%
\bibitem [{\citenamefont {Lebed}\ \emph {et~al.}(2017)\citenamefont {Lebed},
  \citenamefont {Mitchell},\ and\ \citenamefont {Swanson}}]{exotica}%
  \BibitemOpen
  \bibfield  {author} {\bibinfo {author} {\bibfnamefont {R.~F.}\ \bibnamefont
  {Lebed}}, \bibinfo {author} {\bibfnamefont {R.~E.}\ \bibnamefont {Mitchell}},
  \ and\ \bibinfo {author} {\bibfnamefont {E.~S.}\ \bibnamefont {Swanson}},\
  }\href {\doibase 10.1016/j.ppnp.2016.11.003} {\bibfield  {journal} {\bibinfo
  {journal} {Prog. Part. Nucl. Phys.}\ }\textbf {\bibinfo {volume} {93}},\
  \bibinfo {pages} {143} (\bibinfo {year} {2017})},\ \Eprint
  {http://arxiv.org/abs/1610.04528} {arXiv:1610.04528 [hep-ph]} \BibitemShut
  {NoStop}%
\bibitem [{\citenamefont {Chen}\ \emph {et~al.}(2016)\citenamefont {Chen},
  \citenamefont {Chen}, \citenamefont {Liu},\ and\ \citenamefont {Zhu}}]{xyz1}%
  \BibitemOpen
  \bibfield  {author} {\bibinfo {author} {\bibfnamefont {H.-X.}\ \bibnamefont
  {Chen}}, \bibinfo {author} {\bibfnamefont {W.}~\bibnamefont {Chen}}, \bibinfo
  {author} {\bibfnamefont {X.}~\bibnamefont {Liu}}, \ and\ \bibinfo {author}
  {\bibfnamefont {S.-L.}\ \bibnamefont {Zhu}},\ }\href {\doibase
  10.1016/j.physrep.2016.05.004} {\bibfield  {journal} {\bibinfo  {journal}
  {Phys. Rept.}\ }\textbf {\bibinfo {volume} {639}},\ \bibinfo {pages} {1}
  (\bibinfo {year} {2016})},\ \Eprint {http://arxiv.org/abs/1601.02092}
  {arXiv:1601.02092 [hep-ph]} \BibitemShut {NoStop}%
\bibitem [{\citenamefont {Esposito}\ \emph {et~al.}(2015)\citenamefont
  {Esposito}, \citenamefont {Guerrieri}, \citenamefont {Piccinini},
  \citenamefont {Pilloni},\ and\ \citenamefont {Polosa}}]{xyz2}%
  \BibitemOpen
  \bibfield  {author} {\bibinfo {author} {\bibfnamefont {A.}~\bibnamefont
  {Esposito}}, \bibinfo {author} {\bibfnamefont {A.~L.}\ \bibnamefont
  {Guerrieri}}, \bibinfo {author} {\bibfnamefont {F.}~\bibnamefont
  {Piccinini}}, \bibinfo {author} {\bibfnamefont {A.}~\bibnamefont {Pilloni}},
  \ and\ \bibinfo {author} {\bibfnamefont {A.~D.}\ \bibnamefont {Polosa}},\
  }\href {\doibase 10.1142/S0217751X15300021} {\bibfield  {journal} {\bibinfo
  {journal} {Int. J. Mod. Phys. A}\ }\textbf {\bibinfo {volume} {30}},\
  \bibinfo {pages} {1530002} (\bibinfo {year} {2015})},\ \Eprint
  {http://arxiv.org/abs/1411.5997} {arXiv:1411.5997 [hep-ph]} \BibitemShut
  {NoStop}%
\bibitem [{\citenamefont {Giacosa}\ \emph {et~al.}(2019)\citenamefont
  {Giacosa}, \citenamefont {Piotrowska},\ and\ \citenamefont
  {Coito}}]{Giacosa:2019zxw}%
  \BibitemOpen
  \bibfield  {author} {\bibinfo {author} {\bibfnamefont {F.}~\bibnamefont
  {Giacosa}}, \bibinfo {author} {\bibfnamefont {M.}~\bibnamefont {Piotrowska}},
  \ and\ \bibinfo {author} {\bibfnamefont {S.}~\bibnamefont {Coito}},\ }\href
  {\doibase 10.1142/S0217751X19501732} {\bibfield  {journal} {\bibinfo
  {journal} {Int. J. Mod. Phys. A}\ }\textbf {\bibinfo {volume} {34}},\
  \bibinfo {pages} {1950173} (\bibinfo {year} {2019})},\ \Eprint
  {http://arxiv.org/abs/1903.06926} {arXiv:1903.06926 [hep-ph]} \BibitemShut
  {NoStop}%
\bibitem [{\citenamefont {Huang}\ and\ \citenamefont
  {Weldon}(1975)}]{PhysRevD.11.257}%
  \BibitemOpen
  \bibfield  {author} {\bibinfo {author} {\bibfnamefont {K.}~\bibnamefont
  {Huang}}\ and\ \bibinfo {author} {\bibfnamefont {H.~A.}\ \bibnamefont
  {Weldon}},\ }\href {\doibase 10.1103/PhysRevD.11.257} {\bibfield  {journal}
  {\bibinfo  {journal} {Phys. Rev. D}\ }\textbf {\bibinfo {volume} {11}},\
  \bibinfo {pages} {257} (\bibinfo {year} {1975})}\BibitemShut {NoStop}%
\bibitem [{\citenamefont {Nishijima}(1958)}]{PhysRev.111.995}%
  \BibitemOpen
  \bibfield  {author} {\bibinfo {author} {\bibfnamefont {K.}~\bibnamefont
  {Nishijima}},\ }\href {\doibase 10.1103/PhysRev.111.995} {\bibfield
  {journal} {\bibinfo  {journal} {Phys. Rev.}\ }\textbf {\bibinfo {volume}
  {111}},\ \bibinfo {pages} {995} (\bibinfo {year} {1958})}\BibitemShut
  {NoStop}%
\bibitem [{\citenamefont {Machavariani}\ and\ \citenamefont
  {Faessler}(2011)}]{Machavariani_2011}%
  \BibitemOpen
  \bibfield  {author} {\bibinfo {author} {\bibfnamefont {A.~I.}\ \bibnamefont
  {Machavariani}}\ and\ \bibinfo {author} {\bibfnamefont {A.}~\bibnamefont
  {Faessler}},\ }\href {\doibase 10.1088/0954-3899/38/3/035002} {\bibfield
  {journal} {\bibinfo  {journal} {Journal of Physics G: Nuclear and Particle
  Physics}\ }\textbf {\bibinfo {volume} {38}},\ \bibinfo {pages} {035002}
  (\bibinfo {year} {2011})}\BibitemShut {NoStop}%
\bibitem [{\citenamefont {Shankar}(1997)}]{shankar1997effective}%
  \BibitemOpen
  \bibfield  {author} {\bibinfo {author} {\bibfnamefont {R.}~\bibnamefont
  {Shankar}},\ }\href@noop {} {\enquote {\bibinfo {title} {Effective field
  theory in condensed matter physics},}\ } (\bibinfo {year} {1997}),\ \Eprint
  {http://arxiv.org/abs/cond-mat/9703210} {arXiv:cond-mat/9703210 [cond-mat]}
  \BibitemShut {NoStop}%
\bibitem [{\citenamefont {Chew}(1962)}]{Chew:1962mpd}%
  \BibitemOpen
  \bibfield  {author} {\bibinfo {author} {\bibfnamefont {G.~F.}\ \bibnamefont
  {Chew}},\ }\href {\doibase 10.1103/RevModPhys.34.394} {\bibfield  {journal}
  {\bibinfo  {journal} {Rev. Mod. Phys.}\ }\textbf {\bibinfo {volume} {34}},\
  \bibinfo {pages} {394} (\bibinfo {year} {1962})}\BibitemShut {NoStop}%
\bibitem [{\citenamefont {Bazavov}\ \emph
  {et~al.}(2012{\natexlab{b}})\citenamefont {Bazavov} \emph
  {et~al.}}]{Bazavov:2011nk}%
  \BibitemOpen
  \bibfield  {author} {\bibinfo {author} {\bibfnamefont {A.}~\bibnamefont
  {Bazavov}} \emph {et~al.},\ }\href {\doibase 10.1103/PhysRevD.85.054503}
  {\bibfield  {journal} {\bibinfo  {journal} {Phys. Rev. D}\ }\textbf {\bibinfo
  {volume} {85}},\ \bibinfo {pages} {054503} (\bibinfo {year}
  {2012}{\natexlab{b}})},\ \Eprint {http://arxiv.org/abs/1111.1710}
  {arXiv:1111.1710 [hep-lat]} \BibitemShut {NoStop}%
\end{thebibliography}%

\end{document}